\documentclass[epj]{svjour}
\usepackage{dblfloatfix}
\usepackage[belowskip=-5pt,aboveskip=5pt]{caption}

\usepackage{preprintcover}  
\PreprintCoverPaperTitle{Jet energy resolution in proton-proton collisions at $\mathrm{\sqrt{s}=7}$~TeV recorded in 2010 with the ATLAS detector}
\PreprintIdNumber{CERN-PH-EP-2012-191}  
\PreprintCoverAbstract{
The measurement of the jet energy resolution is presented using data recorded with the \ATLAS{} detector in proton-proton collisions at $\sqrt{s}=7$ TeV. The sample corresponds to an integrated luminosity of 35 pb$^{-1}$.  Jets are reconstructed from energy deposits measured by the calorimeters and calibrated using different jet calibration schemes.  The jet energy resolution is measured with two different in situ methods which are found to be in agreement within uncertainties. The total uncertainties on these measurements range from 20\% to 10\% for jets within $|y|<2.8$ and with transverse momenta increasing from 30~GeV to 500~GeV.  Overall, the Monte Carlo simulation of the jet energy resolution agrees with the data within 10\%.}  
\PreprintJournalName{EPJC}  

\RequirePackage[colorlinks,breaklinks]{hyperref}
\hypersetup{linkcolor=blue,citecolor=blue,filecolor=black,urlcolor=blue}

\usepackage{atlasphysics} 
\usepackage{wrapfig}
\usepackage{graphicx}
\usepackage{mathptmx}
\usepackage{amssymb}
\usepackage{helvet}
\usepackage{subfig}
\usepackage{epsfig}
\usepackage{multirow}
\usepackage{rotating}
\usepackage{amsmath}
\usepackage{mathrsfs}
\usepackage{bm}
\usepackage{tikz}
\usetikzlibrary{arrows,3d}
\usepackage[switch]{lineno} 
\usepackage{cite}
\usepackage{url}
\usepackage{fixltx2e} 
\usepackage[mathcal]{eucal}

\newcommand{\alpgen}  {{\sc Alpgen}}
\newcommand{\pythia}  {{\sc Pythia}}
\newcommand{\Perugia} {{\sc Perugia}}
\newcommand{\geant}   {{\sc Geant4}}
\newcommand{\herwig}  {{\sc Herwig}}
\newcommand{\herwigpp}{{\sc Herwig++}}
\newcommand{\jimmy}   {{\sc Jimmy}}
\newcommand{\parp}  {{\sc PARP90}}

\newcommand{\ATLAS}   {\texttt{ATLAS}}

\newcommand{\LHC}    {\texttt{LHC}}

\newcommand{\tabpar}{\hspace{\the\parindent}}

\def\T {\ensuremath{{\rm T}}}

\def\calo {\ensuremath{{\rm calo}}}
\def\part {\ensuremath{{\rm part}}}

\runningpagewiselinenumbers 
\makeatletter
\renewcommand\dbltopfraction{0.75}
\renewcommand\dblfloatpagefraction{0.75}

\makeatother
\usepackage{makeidx}
\makeindex

\begin{document}
\title{Jet energy resolution in proton-proton collisions at $\mathrm{\sqrt{s}=7}$~TeV recorded in 2010 with the ATLAS detector
}
\author{The ATLAS Collaboration}
\date{Received: date / Revised version: date}
\newpage
\abstract{
The measurement of the jet energy resolution is presented using data recorded with the \ATLAS{} detector in proton-proton collisions at $\sqrt{s}=7$ TeV. The sample corresponds to an integrated luminosity of 35 pb$^{-1}$.
Jets are reconstructed from energy deposits measured by the
calorimeters and calibrated using different jet calibration schemes.
The jet energy resolution is measured with two different in situ methods which are found to be in agreement within uncertainties. The total uncertainties on these measurements range from 20\% to 10\% for jets within $|y|<2.8$ and with transverse momenta increasing from 30~GeV to 500~GeV. 
Overall, the Monte Carlo simulation 
of the jet energy resolution agrees with the data within 10\%.
\vskip 2cm}
\authorrunning{\ATLAS{} collaboration} 
\titlerunning{Jet Energy Resolution with the \ATLAS{} detector}
\maketitle
\setcounter{tocdepth}{2}
\tableofcontents
\newpage
\section{Introduction}\label{sec:Intro}
\hspace{\the\parindent}
Precise knowledge of the jet energy resolution is of key importance for the measurement of the cross-sections of inclusive jets, dijets, multijets or vector bosons accompanied by jets~\cite{atlasjet2010,dijetdphi,atlasmultijet2010,atlaswzjets2010}, top-quark cross-sections and mass measurements~\cite{toppaper}, and searches involving resonances decaying to jets~\cite{Atlasdimass,Aad:2011aj}.
The jet energy resolution also has a direct impact on the determination of the missing transverse energy, which plays an important role in many searches for new physics with jets in the final state~\cite{Atlassusy1lepton,susyoleptons}.
This article presents the determination with the \ATLAS{} detector~\cite{ATLAS,TDR} of the jet energy resolution in proton-proton collisions at a centre-of-mass energy of $\sqrt{s}=7$~\TeV{}.  The data sample was collected during 2010 and corresponds to 35 pb$^{-1}$ of integrated luminosity delivered by the Large Hadron Collider (\LHC)~\cite{Evans} at CERN.

The jet energy resolution is determined by exploiting the transverse momentum balance in events containing jets with large transverse momenta (\pt).
This article is structured as follows: Section~\ref{sec:Calorimeter} describes the \ATLAS{} detector. 
Sections~\ref{sec:Simulation},~\ref{sec:EventSelection} and~\ref{sec:JES} respectively introduce the Monte Carlo simulation, the event and jet selection criteria, and the jet calibration methods. 
The two techniques to estimate the jet energy resolution from calorimeter observables, the {\it dijet balance method}~\cite{DZero} and the {\it bisector method}~\cite{UA2}, are discussed respectively in Sections~\ref{sec:dijet} and~\ref{sec:bisector}. 
These methods rely on somewhat different assumptions, which can
be validated in data and are sensitive to different sources of systematic uncertainty. As such, the use of these two independent in situ measurements of the jet energy resolution is important to validate the Monte Carlo simulation. 
Section~\ref{sec:PerfEMJES} presents the results obtained for data and simulation for the default jet energy calibration scheme implemented in \ATLAS{}. 
Section~\ref{sec:closure} compares results of the Monte Carlo simulation in situ methods to the resolutions obtained by comparing the jet energy at calorimeter and particle level. This comparison will be referred to as a closure test.
Sources of systematic uncertainty on the jet energy resolution estimated using the available Monte Carlo simulations and collision data are discussed 
in Section~\ref{sec:systematics}. 
The results for other jet energy calibration schemes are discussed in Sections~\ref{sec:PerfOther} and~\ref{sec:TBJC},
and the conclusions can be found in Section~\ref{sec:Summary}.

\section{The ATLAS detector}\label{sec:Calorimeter}
\hspace{\the\parindent}The \ATLAS{} detector is a multi-purpose detector designed to observe particles produced in high energy proton-proton collisions. A detailed description can be found in Refs.~\cite{ATLAS,TDR}. 
The Inner (tracking) Detector has complete azimuthal coverage and spans the pseudorapidity region $|\eta| < 2.5$ \footnote{The \ATLAS{} reference system is a Cartesian right-handed coordinate system, with the nominal collision point at the origin. The anti-clockwise beam direction defines the positive $z$-axis, with the $x$-axis pointing to the centre of the LHC ring. The angle $\phi$ defines the direction in the plane transverse to the beam ($x,y$). The pseudorapidity is given by $\eta = -\ln\tan\frac{\theta}{2}$, where the polar angle $\theta$ is taken with respect to the positive $z$ direction. The rapidity is defined as $y = 0.5 \times \ln[(E+p_{z})/(E-p_{z})]$, where $E$ denotes the energy and $p_{z}$ is the component of the momentum along the $z$-axis.}. The Inner Detector consists of layers of silicon pixel, silicon microstrip and transition radiation tracking detectors. These sub-detectors are surrounded by a superconducting solenoid that produces a uniform 2 T axial magnetic field.

The calorimeter system is composed of several sub-detectors. A high-granular\-ity liquid-argon (LAr) electromagnetic sampling calorimeter covers the $|\eta| < 3.2$ range, and it is split into a barrel ($|\eta| < 1.475$) and two end-caps ($ 1.375 < |\eta| < 3.2$). Lead absorber plates are used over its full coverage. The hadronic calorimetry in the barrel is provided by a sampling calorimeter using steel as the absorber material and scintillating tiles as active material in the range $|\eta| < 1.7 $. This tile hadronic calorimeter is separated into a large barrel and two smaller extended barrel cylinders, one on either side of the central barrel. In the end-caps, copper/LAr technology is used for the hadronic end-cap calorimeters (HEC), covering the range $1.5 <|\eta| < 3.2$. The copper-tungsten/LAr forward calori\-meters (FCal) provide both electromagnetic and hadronic energy measurements, extending the coverage to $|\eta| = 4.9$.

The trigger system consists of a hardware-based Level 1 (L1) and a two-tier, software-based High Level Trigger (HLT). The L1 jet trigger uses a sliding window algorithm with coarse-granu\-lar\-ity calorimeter towers. This is then refined using jets reconstructed from calorimeter cells in the HLT. 

\section{Monte Carlo simulation}\label{sec:Simulation}

\subsection{Event generators}\label{sec:EventGenerators}
\hspace{\the\parindent}
Data are compared to Monte Carlo (MC) simulations of jets with large transverse momentum produced via strong interactions descri\-bed by Quantum Chromodynamics (QCD) in proton-pro\-ton collisions at a centre-of-mass energy of $\sqrt{s}$ = 7~TeV. The jet energy resolution is derived from several simulation models in order to study its dependence 
on the event generator, on the parton showering and hadronisation models, and on tunes of other soft model parameters, such as those of the undelying event.
The event generators used for this analysis are described below.
\begin{enumerate}
\item \pythia{} 6.4 MC10 tune:
\index{\pythia{}}  \index{ MC10 tune}   
The event generator \pythia{} \cite{pythia}
simulates non-diffractive proton-proton collisions using a $2 \to 2$ matrix element
at the leading order (LO) of the strong coupling constant to model the hard sub-process,
and uses \pt-ordered parton showers to model additional radiation in the leading-logarithm approximation \cite{pythiapartonshower}. 
Multiple parton interactions \cite{Sjostrand:2004ef}, as well as fragmentation and hadronization based on the Lund string model \cite{lundstring} are also simulated. The parton distribution function (PDF)
set used is the modified leading-order MRST LO* set~\cite{PDF-MRST}. The parameters used to describe multiple parton interactions are denoted as the ATLAS MC10 tune~\cite{MC10}. This generator and tune are chosen as the baseline for the jet energy resolution studies.

\index{ \pythia{} \Perugia2010 tune}   
\item The \pythia{} \Perugia2010 tune is an independent tune of \pythia{} to hadron collider data with increased 
final-state radiation to better reproduce the jet and hadronic event shapes observed in LEP and Tevatron data~\cite{Perugia2010}. 
Parameters sensitive to the production of particles with stran\-geness and related to jet fragmentation have also been adjusted. It is the tune favoured by ATLAS jet shape measurements~\cite{Atlasjetshape}.

\index{\pythia \parp{} systematic variation}
\item The \pythia{} \parp{} modification is an independent systematic variation of \pythia{}. The variation has been carried out by changing the parameter that controls the energy dependence of the cut-off, deciding whether the events are generated with the matrix element and parton-shower approach, or the soft underlying event
\cite{MCtunes}.

\index{\pythia8{}}  
\item \pythia{}8 \cite{Pythia8} is based on the event generator \pythia{}
and contains several modelling improvements, such as fully interleaved \pt-ordered evolution of multiparton interactions and initial- and final-state radiation, and a richer mix of underlying-event processes. Once fully tested and tuned, it is expected to offer a complete replacement for version 6.4.

\index{\herwigpp{}}   
\item The \herwigpp{} generator~\cite{herwigpp,herwig,herwig2,herwig3} uses a leading order $2 \to 2$ matrix element with angular-ordered parton showers in the leading-logarithm approximation. Ha\-dronization is performed in the cluster model\cite{herwigclustermodel}.
The underlying event and soft inclusive interactions use 
hard and soft multiple partonic interaction models \cite{HerwigppUI}.
The MRST LO* PDFs \cite{PDF-MRST} are used.

\index{\alpgen{}}   
\item \alpgen{}
is a tree-level matrix element generator for hard multi-parton processes ($2 \to n$) 
in hadronic collisions ~\cite{Alpgen}. 
It is interfaced to \herwig{} to produce parton showers in leading-logarithm 
approximation, which are matched to the matrix element partons with the MLM matching scheme \cite{MLM}. 
\herwig{} is used for hadronization and \jimmy~\cite{jimmy} is used to model soft multiple parton interactions. The LO
CTEQ6L1 PDFs \cite{PDF-CTEQ} are used.
\end{enumerate}

\subsection{Simulation of the ATLAS detector}
\label{sec:GEANT}
\hspace{\the\parindent}Detector simulation is performed with the \ATLAS{} simulation framework~\cite{ATLASsimulation} based on \geant~\cite{GEANT4}, which includes a detailed description of the geometry and the material of the detector. The set of processes that describe hadronic interactions in the \geant~detector simulation are outlined in Refs.~\cite{QGSP,BERT}. The energy deposited by particles in the active detector material is converted into detector signals to mimic the detector read-out. Finally, the Monte Carlo generated events are processed through the trigger simulation of the experiment and are reconstructed and analysed with the same software that is used for data.

\subsection{Simulated pile-up samples}
\label{sec:PileupMCSample}
\hspace{\the\parindent} The nominal MC simulation does not include additional proton-proton interactions (pile-up). In order to study its effect on the jet energy resolution, two additional MC samples are used. The first one simulates additional proton-proton interactions in the same bunch crossing (in-time pile-up) while the second sample in addition simulates effects on calorimeter cell energies from close-by bunches (out-of-time pile-up). The average number of interactions per event is $1.7$ ($1.9$) for the in-time (in-time plus out-of-time) pile-up samples, which is a good representation of the 2010 data.

\section{Event and jet selection}\label{sec:EventSelection}

\hspace{\the\parindent}
The status of each sub-detector and trigger, as well as reconstructed physics objects in \ATLAS{} is continuously assessed by inspection of a standard set of distributions, and data-quality flags are recorded in a database for each luminosity block (of about two minutes of data-taking). This analysis selects events   satisfying data-quality criteria for the Inner Detector and the calorimeters, and for track, jet, and missing transverse energy reconstruction~\cite{JESnote}. 

For each event, the reconstructed primary vertex position is required to be consistent with the beamspot, both transversely and longitudinally, 
and to be reconstructed from at least five tracks with transverse momentum 
$\pt^{\rm{track}} > 150$ MeV associated with it. The primary vertex is defined as the one with the highest associated sum of squared track transverse momenta $\Sigma (\pt^{\rm track})^2$, where the sum runs over all tracks used in the vertex fit. 
Events are selected by requiring a specific OR combination of inclusive single-jet and dijet calorimeter-based triggers~\cite{jetTriggers,jetTriggerCONF}. The combinations are chosen such that the trigger 
efficiency for each $\pt$ bin is greater than 99\%. For the lowest $\pt$ bin (30--40 GeV), this requirement is relaxed,
allowing the lowest-threshold calorimeter inclusive single-jet trigger to be used with an efficiency above 95$\%$.

Jets are reconstructed with the anti-$k_{t}$ jet algorithm~\cite{antiktalg} using the FastJet 
software~\cite{Fastjet} 
with radius parameters $R$~=~0.4 or $R = 0.6$, a four-momentum recombination scheme, and three-dimensional calorimeter topological clusters~\cite{topoClusters} as inputs. Topological clusters are built from calorimeter cells with a signal at least four times higher than the root-mean-square (RMS) of the noise distribution (seed cells). Cells neighbouring the seed which have a signal to RMS-noise ratio $\ge2$ are then iteratively added. Finally, all nearest neighbour cells are added to the cluster without any threshold.

Jets from non-collision backgrounds (e.g.\ beam-gas events) and instrumental noise are removed using the selection criteria outlined in Ref.~\cite{JESnote}.

Jets are categorized according to their reconstructed rapidity in four different regions to account for the differently instrumented parts of the calorimeter: 
\begin{itemize}
\item Central region               ($|y| < 0.8$).
\item Extended Tile Barrel ($0.8 \le |y| < 1.2$).
\item Transition region    ($1.2 \le |y| < 2.1$).
\item End-Cap region       ($2.1 \le |y| < 2.8$).
\end{itemize} 
Events are selected only if the transverse momenta of the two leading jets are above a jet reconstruction threshold of 7~GeV at the electromagnetic scale (see Section~\ref{sec:JES})
and within $|y| \le 2.8$, at least one of them being in the central region. The analysis is restricted to $|y| \le 2.8$ because of the limited number of jets at higher rapidities.

Monte Carlo simulated ``particle jets'' are defined 
as those built using the same jet algorithm as described above, but using instead as inputs the stable particles from the event generator (with a lifetime longer than $10$~ps), excluding muons and neutrinos.

\section{Jet energy calibration}\label{sec:JES}

\hspace{\the\parindent}
Calorimeter jets are reconstructed from calorimeter energy deposits measured at the electromagnetic scale (EM-scale), the baseline signal scale for the energy deposited by electromagnetic showers in the calorimeter. Their transverse momentum is referred to as $\pT^{\rm{EM-scale}}$. For hadrons this leads to a jet energy measurement that is typically 15--55\% 
lower than the true energy,
due mainly to the non-compensating nature of the \ATLAS{} calori\-meter~\cite{CSC}. The jet response is defined as the ratio of calorimeter jet $\pt$ and particle jet $\pt$, reconstructed with the same algorithm, and matched in $\eta-\phi$ space (see Section~\ref{sec:closure}). 

Fluctuations of the hadronic shower, in particular of its electromagnetic content, as well as energy losses in the dead material lead to a degraded resolution and jet energy measurement compared to particles interacting only electromagnetically. Several complementary jet calibration schemes with different levels of complexity and different sensitivity to systematic effects have been developed to understand the jet energy measurements. 
The jet calibration is performed by applying corrections derived from Monte Carlo simulations to restore the jet response to unity. This is referred to as determining the jet energy scale (JES).

The analysis presented in this article aims to determine the jet energy resolution for jets reconstructed using various JES strategies. A simple calibration, referred to as the EM+JES calibration scheme, has been chosen for the 2010 data~\cite{JESnote}. 
It allows a direct evaluation of the systematic uncertainties from single-hadron response measurements and is therefore suitable for first physics analyses. More sophisticated calibration techniques to improve the jet resolution and reduce partonic flavour response 
differences have also been developed.
They are the Local Cluster 
Weighting (LCW), the Global Cell Weighting (GCW) and the Global Sequential (GS) methods~\cite{JESnote}. In addition to these calorimeter calibration schemes, a Track-Based Jet Correction (TBJC) has been derived to adjust the response and reduce fluctuations on a jet-by-jet basis without changing the average jet energy scale. These calibration techniques are briefly described below.  

\subsection{The EM+JES calibration}

\hspace{\the\parindent}For the analysis of the first proton-proton collisions, a simple Monte Carlo simulation-based correction is applied as the default to restore the hadronic energy scale on average. The EM+JES calibration scheme applies corrections as a function of the jet transverse momentum and pseudorapidity to jets reconstructed at the electromagnetic scale. The main advantage of this approach is that 
it allows the most direct evaluation of the systematic uncertainties.
The uncertainty on the absolute jet energy scale was determined to be less than $\pm 2.5\% $ in the central calorimeter region ($|y| < 0.8$) and $\pm 14\% $ in the most forward region ($3.2 \le |y| < 4.5$) for jets with $\pt > 30$~GeV~\cite{JESnote}. These uncertainties were evaluated using test-beam results, single hadron response in situ measurements, comparison with jets built from tracks, $\pt$ balance in dijet and $\gamma$+jet events, estimations of pile-up energy deposits, and detailed Monte Carlo comparisons.

\subsection{The Local Cluster Weighting (LCW) calibration}

\hspace{\the\parindent}The LCW calibration scheme uses properties of clusters to calibrate them individually {\it{prior}} to jet finding and reconstruction. The calibration weights are determined from Monte Carlo simulations of charged and neutral pions according to the cluster topology measured in the calorimeter. The cluster properties used are the energy density in the cells forming them, the fraction of their energy deposited in the different calorimeter layers, the cluster isolation and its depth in the calorimeter. 
Corrections are applied to the cluster energy to account for the energy deposited in the calorimeter but outside of clusters and energy deposited in material before and in between the calorimeters. Jets are formed from calibrated clusters, and a final correction is applied to the jet energy to account for jet-level effects. The resulting jet energy calibration is denoted as LCW+JES.

\subsection{The Global Cell Weighting (GCW) calibration}

\hspace{\the\parindent}The GCW calibration scheme attempts to compensate for the different calorimeter response to hadronic and electromagnetic energy deposits at cell level. The hadronic signal is characterized by low cell energy densities and, thus, a positive weight is applied. The weights, which depend on the cell energy density and the calorimeter layer only, are determined by minimizing the jet resolution evaluated by comparing reconstructed and particle jets in Monte Carlo simulation. They correct for several effects at once (calorimeter non-compensation, dead material, etc.). A jet-level correction is applied to jets reconstructed from weighted cells to account for global effects. The resulting jet energy calibration is denoted as GCW+JES.

\subsection{The Global Sequential (GS) calibration}

\hspace{\the\parindent}The GS calibration scheme uses the longitudinal and transverse structure of the jet calorimeter shower to compensate for fluctuations in the jet energy measurement. In this scheme the jet energy response is first calibrated with the EM+JES calibration. Subsequently, the jet properties are used to exploit the topology of the energy deposits in the calorimeter to characterize fluctuations in the hadronic shower development.  These corrections are applied such that the mean jet energy is left unchanged, and each correction is applied sequentially. This calibration is designed to improve the jet energy resolution without changing the average jet energy scale.

\subsection{Track-based correction to the jet calibration}\label{sec:TBJCintro}

\hspace{\the\parindent}
Regardless of the inputs, algorithms and calibration methods chosen for
calorimeter jets, 
more information on the jet topology can be obtained 
from reconstructed tracks associated to the jet.
Calibrated jets have an average energy
response close to unity. 
However, the energy of an individual jet can be over- or underestimated depending on several factors, for example: the ratio of the electromagnetic and hadronic components of the jet; the fraction of energy lost in dead material, in either the inner detector, the solenoid, the cryostat before the LAr, or the cryostat between the LAr and the TileCal.
The reconstructed tracks associated to the jet are sensitive to some of these effects and therefore 
can be used to correct the calibration on a jet-by-jet basis. 

In the method referred to as Track-Based Jet Correction (TBJC)~\cite{CSC}, 
the response is adjusted depending on the number of tracks associated with
the jet.  The jet energy response is observed to decrease with jet track
multiplicity mainly 
because the ratio of the electromagnetic to the hadronic component decreases on average as the number of tracks increases.  
In effect, a low charged-track multiplicity typically indicates a
predominance of neutral hadrons, in particular $\pi^0$s which yield
electromagnetic deposits in the calorimeter with $R\simeq1$.  A large number of
charged particles, on the contrary, signals 
a more dominant hadronic component,
with a lower response due to the non-compensating nature of the
calorimeter ($h/e<1$). 
The TBJC method is designed to be applied as an option in addition to any JES calibration scheme, since it does not change the overall response, to reduce the jet-to-jet energy fluctuations and improve the resolution.

\section{In situ jet resolution measurement using the dijet balance method}\label{sec:dijet}

Two methods are used in dijet events to measure in situ the fractional jet $\pt$ resolution, $\sigma(\pt)/\pt$, which at fixed rapidity is equivalent to the fractional jet energy resolution, $\sigma(E)/E$.
The first method, presented in this section, relies on the approximate scalar balance between the transverse momenta of the two leading jets and measures the sensitivity of this balance to the presence of extra jets directly from data. The second one, presented in the next section, uses the projection of the vector sum of the leading jets' transverse momenta on the coordinate system bisector of the azimuthal angle between the transverse momentum vectors of the two jets. It takes advantage of the very different sensitivities of each of these projections to the underlying physics of the dijet system and to the jet energy resolution.  

\subsection{Measurement of resolution from asymmetry}

\hspace{\the\parindent}The dijet balance method for the determination of the jet \pt\ resolution is based on momentum conservation in the transverse plane.
The asymmetry between the transverse momenta of the two leading jets $A(p_{\T,1},p_{\T,2})$ is defined as
\begin{equation}
A(p_{\T,1},p_{\T,2}) \equiv \frac{p_{\T,1}-p_{\T,2}}{p_{\T,1}+p_{\T,2}} .
\label{Asym}
\end{equation}
where $p_{\T,1}$ and $p_{\T,2}$ refer to the randomly ordered transverse momenta of the two leading jets. The width $\sigma(A)$ of a Gaussian fit to $A(p_{\T,1},p_{\T,2})$ is used to characterize the asymmetry distribution and determine the jet \pt~resolutions. 

For events with exactly two particle jets that satisfy the hypothesis of momentum balance in the transverse plane,
and requiring both jets to be in the same rapidity region, the relation between $\sigma(A)$ and the fractional jet resolution is given by
\begin{equation}
\sigma(A) = \frac{\sqrt{{\sigma^2(p_{\T,1}) + \sigma^2(p_{\T,2})}}}{\langle p_{\T,1} + p_{\T,2} \rangle} \simeq
\frac{1}{\sqrt{2}} \frac{\sigma(\pt)}{\pt}, 
\label{eq:dijet_formula}
\end{equation}
where 
$\sigma(p_{\T,1}) = \sigma(p_{\T,2}) = \sigma(p_\T)$, since both jets are in the same $y$ region.

If one of the two leading jets ($j$) is in the rapidity bin being probed and the other one  ($i$)  in a reference $y$ region, it can be shown that the fractional jet \pt~resolution is given by
\begin{equation}
\frac{\sigma(p_\T)}{p_\T} \Big{\vert}_{(j)} = \sqrt{4\,\sigma^{2}(A_{(i,j)}) - 2\,\sigma^{2}(A_{(i)})} \ , 
\label{eq:dijet_formula_ij}
\end{equation}
where $A_{(i,j)}$ is measured in a topology with the two jets in different rapidity regions and where $(i) \equiv (i,i)$ denotes both jets in the same $y$ region.  

\begin{figure}[h!]
\centering
\includegraphics[width=1\linewidth]{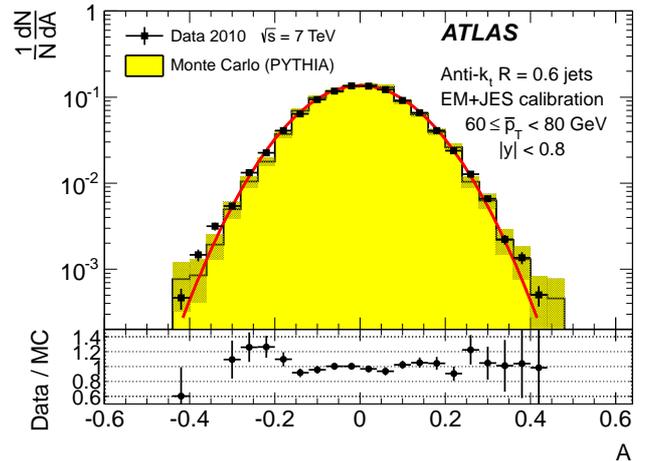}
\caption{Asymmetry distribution as defined in Equation~\eqref{Asym} for $\bar{p}_\T=60-80$~GeV and $|y| < 0.8$. Data (points with error bars) and Monte Carlo simulation (histogram with shaded bands)  are overlaid, together with a Gaussian fit to the data. The lower panel shows the ratio between data and MC simulation.
The errors shown are only statistical.}
\label{fig:Asymmetries}
\end{figure}

The back-to-back requirement is approximated 
by an azimuthal angle cut between the leading jets, $\Delta \phi(j_{1},j_{2}) \geq 2.8$, and a veto on the third jet momentum, $p^{\rm{EM-scale}}_{\T,3} < 10\,\text{GeV}$, with no rapidity restriction.
The resulting asymmetry distribution is shown in Fig.~\ref{fig:Asymmetries} for a $\bar{p}_\T \equiv (p_{\T,1} + p_{\T,2})/2$ bin of $60\,\text{GeV}\le\bar{p}_\T <80\,\text{GeV}$, in the central region ($|y| < 0.8$).  Reasonable agreement in the bulk is observed between data and Monte Carlo simulation.

\subsection{Soft radiation correction}\label{sec:softradiation}
\hspace{\the\parindent}
Although requirements on the azimuthal angle between the leading jets and on the third jet transverse momentum are designed to enrich the purity of the back-to-back jet sample, it is important to account for the presence of additional soft particle jets not detected in the calorimeter.

\begin{figure}[h!]
\centering
\includegraphics[width=1\linewidth]{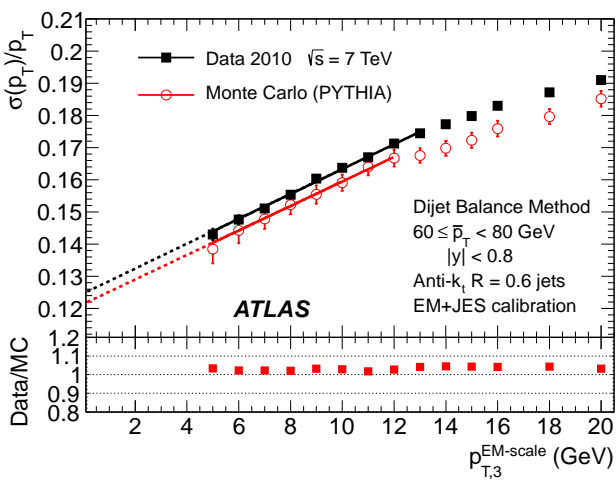}\\
\caption{Fractional jet $\pt$ resolutions, from Equation~\ref{eq:dijet_formula}, measured in events with $60 \le\bar{p}_\T <80$~GeV and with third jet with $\pt$ less than $p_{\T,3}^{\rm{EM-scale}}$, as a function of $p_{\T,3}^{\rm{EM-scale}}$, for data (squares) and Monte Carlo simulation (circles). The solid lines correspond to linear fits while the dashed lines show the extrapolations to $p_{\T,3}^{\rm{EM-scale}}= 0$. 
The lower panel shows the ratio between data and MC simulation.
The errors shown are only statistical.}
\label{fig:SoftRadiationExtrap}
\end{figure}

In order to estimate the value of the asymmetry for a pure particle dijet event, $\sigma(\pt)/\pt \equiv \sqrt{2}\, \sigma(A)$ is recomputed allowing for the presence of an additional third jet in the sample for a series of $p_{\T,3}^{\rm{EM-scale}}$ cut-off threshold values up to 20~GeV. The cut on the third jet
is placed at the EM-scale to be independent of calibration effects and to have a stable reference for all calibration schemes. For each $p_\T$ bin, the jet energy resolutions obtained with the different $p_{\T,3}^{\rm{EM-scale}}$ cuts are fitted with a straight line and extrapolated to $p_{\T,3}^{\rm{EM-scale}}\rightarrow 0$, in order to estimate the expected resolution for an ideal dijet topology
\begin{equation*}
\left.\frac{\sigma(p_\T)}{p_\T}\right|_{p_{\T,3}^{\rm{EM-scale}}\ \rightarrow \  0}.
\end{equation*}
The dependence of the jet $\pt$ resolution on the presence of a third jet is illustrated in Fig.~\ref{fig:SoftRadiationExtrap}. The linear fits and their extrapolations for a $\bar{p}_\T$ bin of $60 \le\bar{p}_\T <80$~GeV are shown. Note that the resolutions become systematically broader
as the $p_{\T,3}^{\rm{EM-scale}}$ cut increases. This is a clear indication that the jet resolution determined from two-jet topologies depends on the presence of additional radiation and on the underlying event.

A soft radiation (SR) correction factor, $K_\mathrm{soft}(\bar{p}_\T)$, is obtained from the ratio 
of the values of the linear fit at 0~GeV and at~10 GeV:
\begin{equation}
K_\mathrm{soft}(\bar{p}_\T) = \frac
{\left.\frac{\sigma(\pt)}{\pt}\right|_{p_{\T,3}^{\rm{EM-scale}} \longrightarrow\; 0{\rm ~GeV}}}
{\left.\frac{\sigma(\pt)}{\pt}\right|_{p_{\T,3}^{\rm{EM-scale}}\,=\,10{\rm ~GeV}}} \ .
\end{equation}
\hspace{\the\parindent}This multiplicative correction is applied to the resolutions extracted from the dijet asymmetry for $p^{\rm{EM-scale}}_{\T,3}<10$~GeV events. The correction varies from 25\% for events with $\bar{p}_\T$~of 50~GeV down to 5\% for $\bar{p}_\T$~of 400~GeV. In order to limit the statistical fluctuations, $K_\mathrm{soft}(\bar{p}_\T)$ is fit with a parameterization of the form $K_\mathrm{soft}(\bar{p}_\T) = a + b/ \,(\log\bar{p}_\T)^2$, which was found to describe the distribution well, within uncertainties. 
The differences in the resolution due to other parameterizations were studied and treated as a systematic uncertainty, resulting in a relative uncertainty of about 6\% (see Section~\ref{sec:systematics}).

\subsection{Particle balance correction}
\begin{figure}[h!]
\centering
\includegraphics[width=1\linewidth,viewport=20 5 550 415,clip]{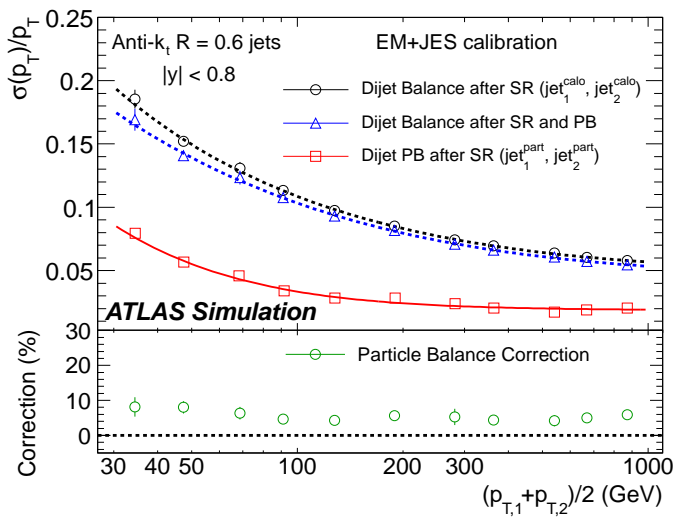}
\vspace{-1mm}
\caption{Fractional jet resolution obtained in simulation using the dijet balance method, shown as a function of $\bar{p}_\T$, both before (circles) and after the particle-balance (PB) correction (triangles). Also shown is the dijet PB correction itself (squares) and, in the lower panel, its relative size with respect to the fractional jet resolution. The errors shown are only statistical.}
\label{fig:PI}
\end{figure}

\hspace{\the\parindent}The $\pt$ difference between the two calorimeter jets is not solely due to resolution effects, but also to the 
balance 
between the respective particle jets,
\begin{equation*}
p_{\T,2}^{\calo}-p_{\T,1}^{\calo} = (p_{\T,2}^{\calo}-p_{\T,2}^{\part}) - (p_{\T,1}^{\calo}-p_{\T,1}^{\part}) + (p_{\T,2}^{\part}-p_{\T,1}^{\part}) .
\end{equation*}
\hspace{\the\parindent}The measured difference (left side) is decomposed into resolution fluctuations (the first two terms on the right side) plus a particle-level 
balance (PB) 
term that originates 
from out-of-jet showering in the particle jets and from soft QCD effects.
In order to correct for this contribution, the particle-level 
balance 
is estimated using the same technique (asymmetry plus soft radiation correction) as for calorimeter jets. 
The contribution of the dijet 
PB
after the SR correction is subtracted in quadrature from the in situ resolution for both data and Monte Carlo simulation. The result of this procedure is shown for simulated events in the central region in Fig.~\ref{fig:PI}.
The relative size of the particle-level 
balance 
correction with respect to the measured resolutions varies between 2\% and 10\%.

\section{In situ jet resolution measurement using the bisector method}\label{sec:bisector}

\subsection{Bisector rationale}
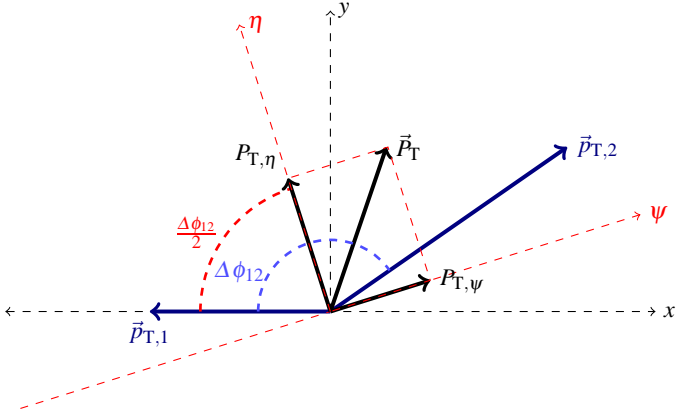
\begin{figure}[h!]
\centering
\begin{tikzpicture}[scale=0.95]
\pgfmathsetmacro{\ejes}{3}
\pgfmathsetmacro{\angu}{145}

\draw [dashed, <->] (-1.5*\ejes,0) -- (1.5*\ejes,0) node[right] {$x$}; 
\draw [dashed, ->] (0,0) -- (0,1.4*\ejes) node[right] {$y$}; 


\draw [line width=1.5,blue!50!black,<-](-2.5,0)node[below] {$\vec{p}_{\T,1}$} -- (0,0); 
\draw [line width=1.5,blue!50!black,->](0,0) -- ++(180-\angu:4)node[right] {$\vec{p}_{\T,2}$}; 

\draw [line width=1.5,black,->](0,0) -- +(-2.5+4*0.819,4*0.574)node[right] {$\vec{P}_\T$}; 

\draw [dashed,line width=1,blue!70](-1,0) arc (180:180-\angu:1);
\node [blue!70] at (180:1) [xshift=-7pt, yshift=15] {$\Delta \phi_{12}$};
\draw [dashed,line width=1,red](-1.8,0) arc (180:180-0.5*\angu:1.8);
\node [red] at (180:1.3) [xshift=-15, yshift=30] {$\frac{\Delta \phi_{12}}{2}$};

\begin{scope}[rotate = 90-0.5*\angu]
  \draw [dashed,red, ->] (-1.5*\ejes,0) -- (1.5*\ejes,0) node[right] {$\psi$}; 
  \draw [dashed,red, ->] (0,0) -- (0,1.4*\ejes) node[right] {$\eta$}; 
  \draw [line width = 1.5,black, ->] (0,0) -- (1.45,0) node[yshift=8pt, below right] {$P_{\T,\psi}$}; 
  \draw [line width = 1.5,black, ->] (0,0) -- (0,1.95) node[above left] {$P_{\T,\eta}$}; 
  \draw [dashed,red] (0,0) rectangle (1.45,1.95);
\end{scope}

\end{tikzpicture} 
\caption{Variables used in the bisector method. The $\eta$-axis corresponds to the azimuthal angular bisector of the dijet system in the plane transverse to the beam, while the $\psi$-axis is defined as the one orthogonal to the $\eta$-axis.}
\label{fig:Bi-Sector}
\end{figure}

\hspace{\the\parindent}The bisector method~\cite{UA2} is based on a transverse 
balance 
vector, $\vec{P}_\T$, defined as the vector sum of the momenta of the two leading jets in
dijet events. This vector is projected along an orthogonal coordinate
system in the transverse plane, ($\psi,\, \eta$), where $\eta$ is
chosen in the direction that bisects the angle formed by $\vec{p}_{\T,1}$ and
$\vec{p}_{\T,2}$, $\Delta\phi_{12} = \phi_{1} -
\phi_{2}$. This is illustrated in Fig.~\ref{fig:Bi-Sector}. 

For a perfectly balanced dijet event, $\vec{P}_{\T}=0$. There are of course a number of sources that give rise to significant fluctuations around this value, and thus to a non-zero variance of its $\psi$ and $\eta$ components, denoted $\sigma_\psi^2 $
and $\sigma_\eta^2$,
respectively.
At particle level,
$\vec{P}^{\part}_\T$ receives contributions mostly from initial-state radiation. This effect is expected to be isotropic in the ($\psi,\eta$) plane, leading to similar fluctuations in both components,
$\sigma^{\part}_{\psi}=\sigma^{\part}_{\eta}$.

The validity of this assumption, which is at the root of the bisector method,  can be checked with Monte Carlo simulations and with data. The precision with which it can be assessed is considered as a systematic uncertainty (see Section~\ref{sec:Validation}). The $\psi$ component has greater sensitivity to the energy resolution because $P_{\T,\psi}$ is the difference between two large transverse momentum components while $P_{\T,\eta}$ is the sum of two small components. Effects such as contamination from 3-jet events or final-state radiation not absorbed in the leading jets by the clustering algorithm could give rise to a $\sigma^{\part}_{\psi}>\sigma^{\part}_{\eta}$.
At calorimeter level, $\sigma^{2\ \calo}_{\psi}$ is expected to be significantly larger than $\sigma^{2\ \calo}_{\eta}$, mostly because of the jet energy resolution.

If both jets belong to the same $y$ region, such that they have the same average jet energy resolution, it can be shown that 
\begin{eqnarray}
\frac{\sigma(\pt)}{\pt} \,=\, 
\frac{\sqrt{\sigma^{2\ \calo}_\psi -\sigma^{2\ \calo}_\eta}}
{\sqrt{2}\, \pt \, \sqrt{\langle|\cos\Delta\phi_{12}|\rangle}} .
\label{eq:bisector_formula}
\end{eqnarray}
The resolution is thus expressed in terms of calorimeter observables only. 
The contribution from soft radiation and the underlying event is minimised by subtracting in quadrature $\sigma_{\eta}$ from  $\sigma_{\psi}$.

If one of the leading jets ($j$) belongs to the rapidity region being probed, and the other one ($i$) to a previously measured reference $y$ region, then
\begin{equation}
\frac{\sigma(\pt)}{\pt}\Big{\lvert}_{(j)} \,=\, 
\sqrt{\frac{\sigma^{2\ \calo}_\psi -\sigma^{2\ \calo}_\eta}
{\ptsq  \, \langle|\cos\Delta\phi_{12}|\rangle}\Big{\lvert}_{(i,j)} - \frac{\sigma^{2}(\pt)}{ \ptsq }\Big{\lvert}_{(i)}}.
\label{eq:bisector_ij}
\end{equation}

\begin{figure}[h!]
\vspace{-2mm}
\centering
\includegraphics[width=1\linewidth]{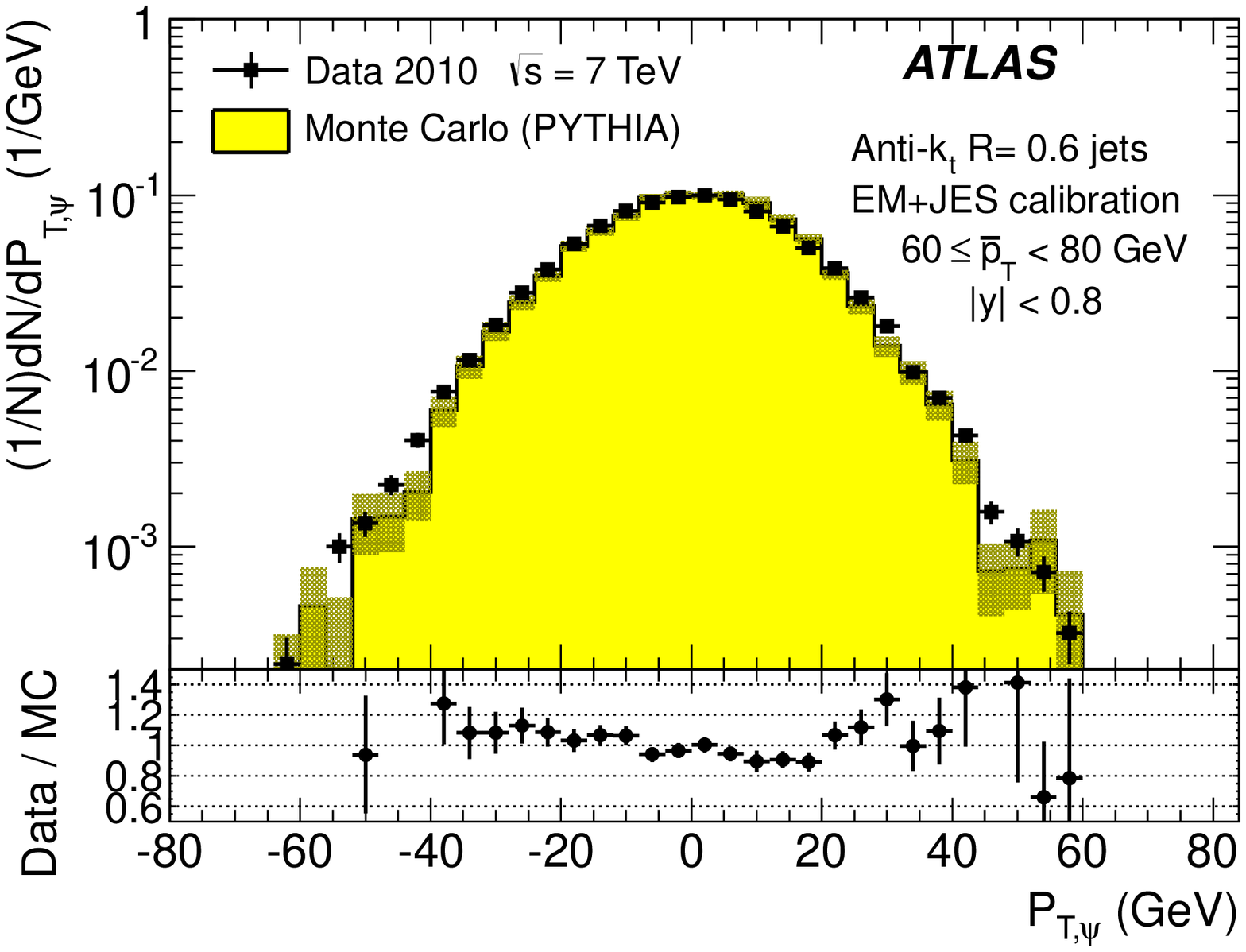}\\
\includegraphics[width=1\linewidth]{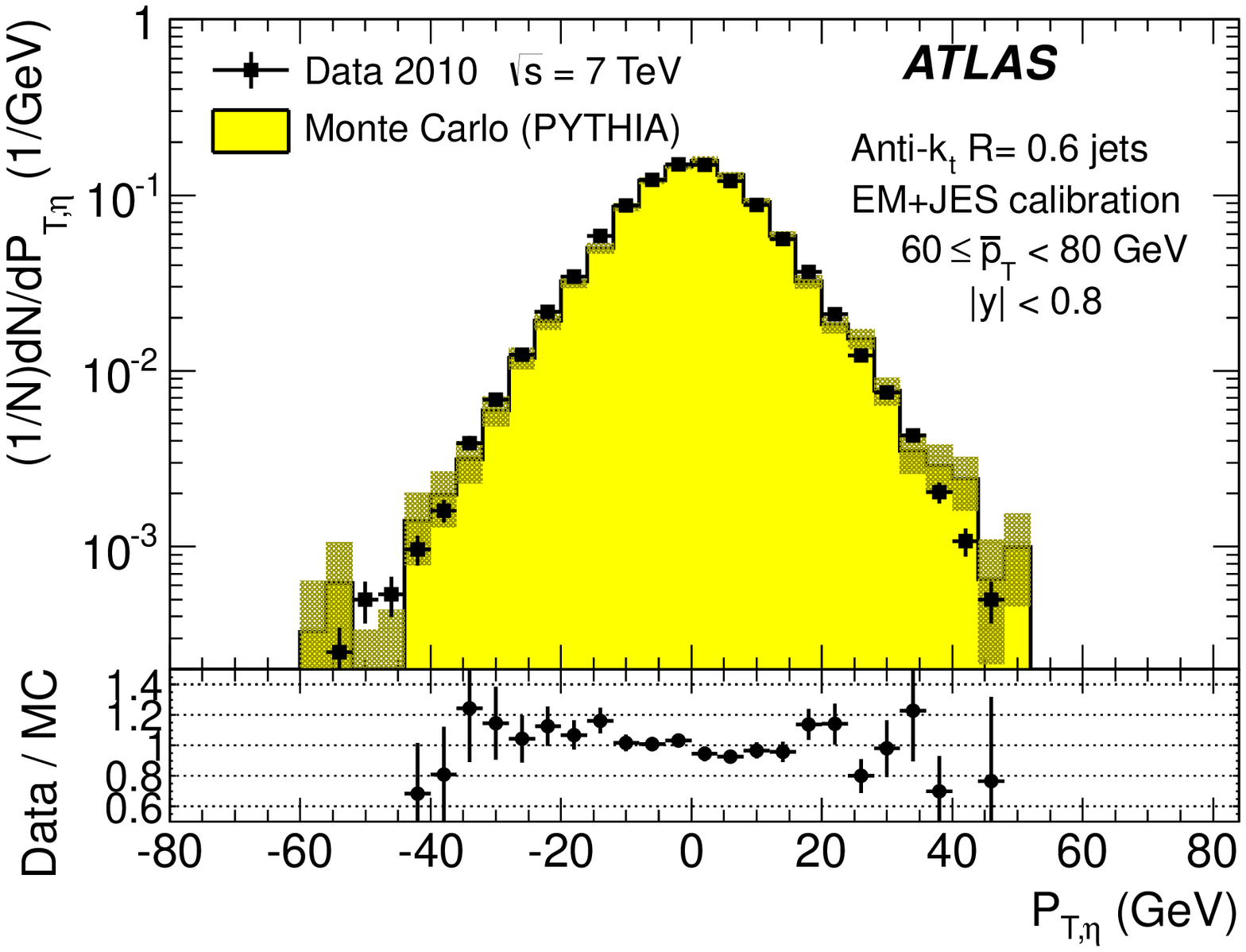}
\vspace{-6mm}
\caption{Distributions of the $P_{\T,\psi}$ (top) and $P_{\T,\eta}$ (bottom) components of the 
balance 
vector $\vec{P}_\T$, for $\bar{p}_\T= 60-80$~GeV. The data (points with error bars) and Monte Carlo simulation (histogram with shaded bands) are overlaid.
The lower panel shows the ratio between data and MC simulation.
The errors shown are only statistical.}
\label{fig:psi_eta}
\end{figure}

\begin{figure}[h!]
\centering
\includegraphics[width=1\linewidth,viewport=0 10 568 410,clip]{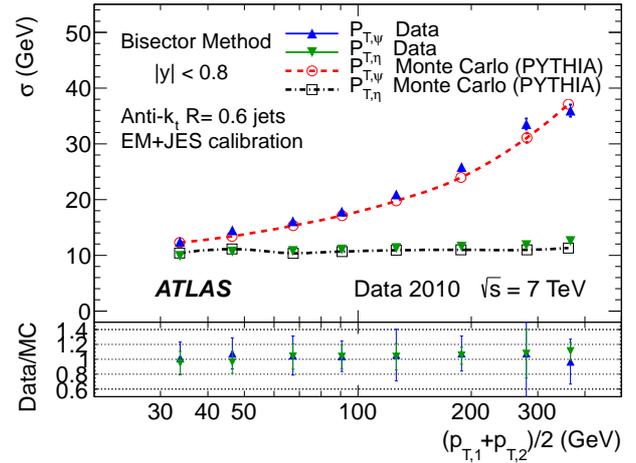}
\caption{Standard deviations of $P_{\T,\psi}$ and $P_{\T,\eta}$, the components of the 
balance 
vector, as a function of $\bar{p}_\T$. 
The lower panel shows the ratio between data and MC simulation.
The errors shown are only statistical.} 
\label{fig:psi_eta_all}
\end{figure}

The dispersions $\sigma_\psi$ and $\sigma_\eta$ are extracted from Gaussian fits to the $P_{\T,\psi}$ and $P_{\T,\eta}$ distributions in bins of $\bar{p}_\T$. There is no $\Delta \phi$ cut imposed between the leading jets, but it is implicitly limited by a $p_{\T,3}^\mathrm{EM-scale}<10$~GeV requirement on the third jet, as discussed in the next section.  Figure~\ref{fig:psi_eta} compares the distributions of $P_{\T,\psi}$ and $P_{\T,\eta}$ between data and Monte Carlo simulation in the momentum bin $60\le\bar{p}_\T<80$~GeV. The distributions agree within statistical fluctuations. 
The resolutions obtained from the $P_{\T,\psi}$ and $P_{\T,\eta}$ components of the 
balance 
vector are summarised in the central region as a function of $\bar{p}_\T$ in Fig.~\ref{fig:psi_eta_all}. As expected, the resolution on the $\eta$ component does not vary with the jet $\pt$, while the resolution on the $\psi$ component degrades as the jet $\pt$ increases.

\subsection{Validation of the soft radiation isotropy with data}\label{sec:Validation}
\begin{figure}[h!]
\centering
\includegraphics[width=1\linewidth,viewport=0 10 568 410,clip]{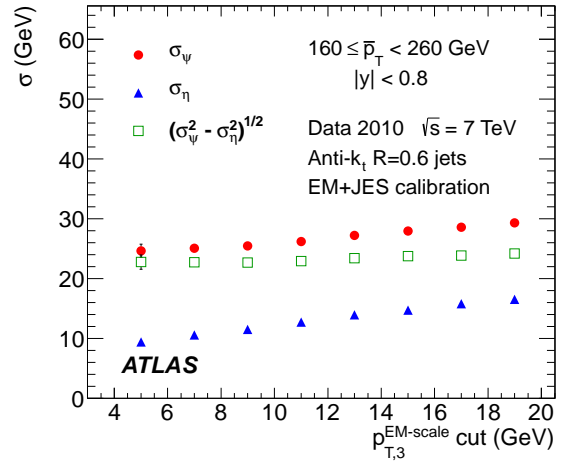}\\
\caption{Standard deviations $\sigma^\calo_\psi$, $\sigma^\calo_\eta$ and $[(\sigma^2_\psi-\sigma^2_\eta)^\calo]^{1/2}$ as a function of the upper $p_{\T,3}^\mathrm{EM-scale}$ cut, for $R = 0.6$  anti-$k_t$ jets with $\bar{p}_\T=160-260$~GeV. 
The errors shown are only statistical.
}
\label{SRval}
\end{figure}

\hspace{\the\parindent}
Figure~\ref{SRval} shows the width of the $\psi$ and $\eta$ components of $\vec{P}_{\T}$ as a function of the $p_{\T,3}^\mathrm{EM-scale}$ cut, for anti-$k_t$ jets with $R = 0.6$.
The two leading jets are required to be in the same rapidity region, $|y|<0.8$, while there is no rapidity restriction for the third jet. As expected, both components increase due to the contribution from soft radiation as the $p_{\T,3}$ cut is increased. Also shown as a function of the $p_{\T,3}^\mathrm{EM-scale}$ cut is the square-root of the difference between their variances, which yields the fractional momentum resolution when divided by $2\,\langle \ptsq\rangle \langle \cos\Delta\phi \rangle$. 

It is observed that the increase of the soft radiation contribution to $\smash[t]{\sigma^\calo_\psi}$ and $\smash[t]{\sigma^\calo_\eta}$ cancels in the squared difference and that it remains almost constant, within statistical uncertainties, up to $p_{\T,3}^\mathrm{EM-scale}\simeq$ 20~GeV for $\bar{p}_\T$ between 160-260~GeV. The same behaviour is observed for other $\bar{p}_\T$ ranges. 
This cancellation demonstrates that the isotropy assumption used for the bisector method is valid over a wide range of choices of $p_{\T,3}^\mathrm{EM-scale}$  without the need for requiring an explicit
$\Delta\phi$ cut between the leading jets. 
The precision with which it can be ascertained in situ that $\sigma^\part_\psi=\sigma^\part_\eta$ is taken conservatively as a systematic uncertainty on the method, of about $4 - 5\%$ at 50~GeV (see Section~\ref{sec:systematics}).

\section{Performance for the EM+JES calibration}\label{sec:PerfEMJES}
\begin{figure}[h!]
\centering
\includegraphics[width=1\linewidth,viewport=35 0 550 325,clip]{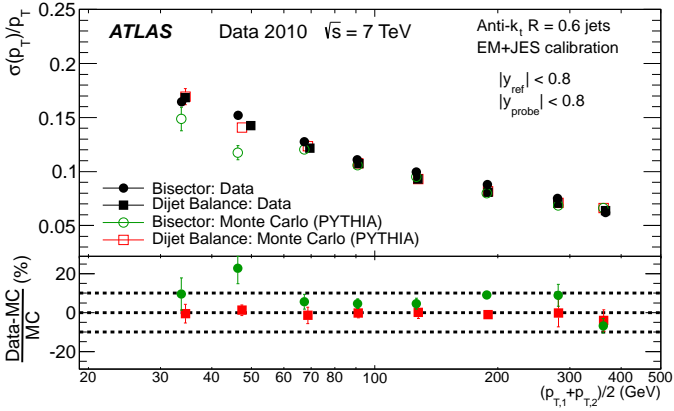}
\caption{Fractional jet \pt\ resolution for the dijet balance and bisector methods as a function of $\bar{p}_\T$. The lower panel shows the relative difference between data and Monte Carlo results. The dotted lines indicate a relative difference of $\pm$10\%. Both methods are found to be in agreement within 10\%  between data and Monte Carlo simulation.
The errors shown are only statistical.} 
\label{fig:Comparison}
\end{figure}

\hspace{\the\parindent}
The performances of the dijet balance and bisector methods are compared for both data and Monte Carlo simulation as a function of jet $\pt$ for jets reconstructed in the central region with the anti-$k_t$ algorithm with $R = 0.6$ and using the EM+JES calibration scheme. The results are shown in Fig.~\ref{fig:Comparison}. The resolutions obtained from the two independent in situ methods are in good agreement with each other within the statistical uncertainties. The agreement between data and Monte Carlo simulation is also good with some deviations observed at low $\pt$.

\begin{figure}[h!]
\centering
\includegraphics[width=1\linewidth,viewport=35 0 550 325,clip]{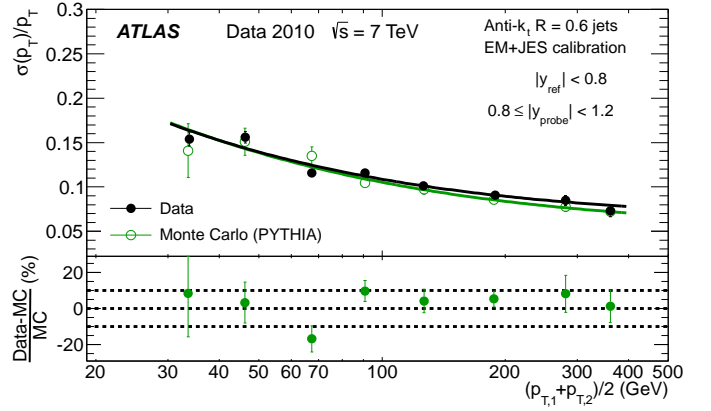}\\
\includegraphics[width=1\linewidth,viewport=35 0 550 325,clip]{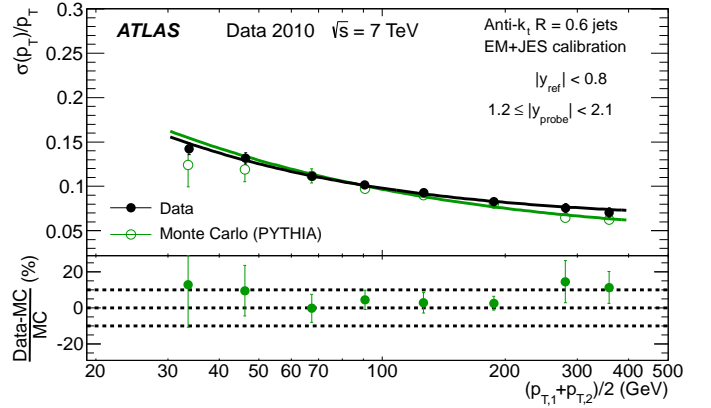}\\
\includegraphics[width=1\linewidth,viewport=35 0 550 325,clip]{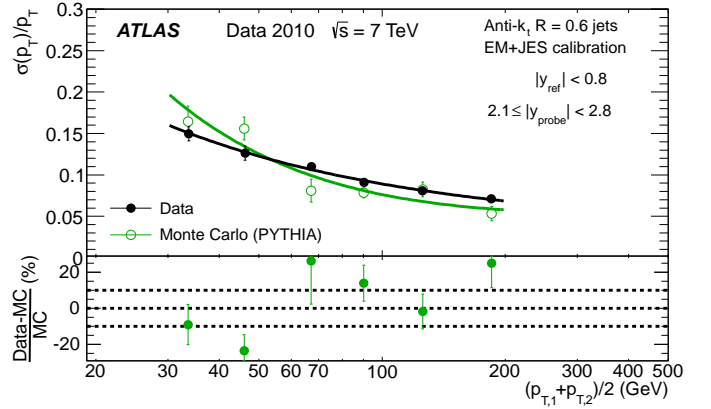}
\caption{Fractional jet \pt\ resolution as a function of $\bar{p}_\T$ for anti-$k_t$ with $R=0.6$ jets in the Extended Tile Barrel (top), Transition (center) and End-Cap (bottom) regions using the bisector method. In the lower panel of each figure, the relative difference between the data and the MC simulation results is shown. The dotted lines indicate a relative difference of $\pm$10\%.
The errors shown are only statistical.
}
\label{fig:ComparisonDiffRegions}
\end{figure}

The resolutions for the three jet rapidity bins with $|y|>0.8$, the Extended Tile Barrel, the Transition and the End-Cap regions, are measured using Eqs.~\ref{eq:dijet_formula_ij} and~\ref{eq:bisector_ij}, taking the central region as the reference. The results for the bisector method are shown in Fig.~\ref{fig:ComparisonDiffRegions}. 
Within statistical errors the resolutions obtained for data and Monte Carlo simulation are in agreement within $\pm10$\% over most of the $\pt$-range in the various regions.

Figure~\ref{fig:ComparisonDiffRegions} shows that dependences are well described by fits to the standard functional form expected for calorimeter-based resolutions, with three independent contributions, the effective noise ($N$), stochastic ($S$) and constant ($C$) terms.

\begin{equation}
\frac{\sigma(\pt)}{\pt} = \frac{N}{\pt}\oplus\frac{S}{\sqrt{\pt}}\oplus\,C\,.
\label{eq:res}
\end{equation}
The $N$ term is due to external noise contributions that are not (or only weakly) dependent on the jet \pt, and include the electronics and detector noise, and contributions from pile-up. It is expected to be significant in the low-$\pt$ region, below $\sim$30~GeV. The $C$ term encompasses the fluctuations that are a constant fraction of the jet \pt, 
assumed at this early stage of data-taking to be due to real signal lost in passive material (e.g.\ cryostats and solenoid coil), to non-uniformities of response across the calorimeter, etc.
It is expected to dominate the high-$\pt$ region, above 400~GeV. 
For intermediate values of the jet $\pt$, the statistical fluctuations, 
represented by the $S$ term, become the limiting factor in the resolution. With the present data sample that covers a restricted $\pt$ range, 30~GeV $\leq\pt<$ 500~GeV, there is a high degree of correlation between the fitted parameters and it is not possible to unequivocally disentangle their contributions.
\section{Closure test using Monte Carlo simulation}\label{sec:closure}
\hspace{\the\parindent}
The Monte Carlo simulation expected resolution is derived 
considering matched particle and calorimeter jets in the event, with no back-to-back geometry requirements. Matching is done in $\eta$ -- $\phi$ space, and jets are associated if $\Delta R = \sqrt{(\Delta\eta)^2 +(\Delta\phi)^2 }$ $< 0.3 $. The jet response is defined as $\pt^{\rm calo} / \pt^{\rm part}$, in bins of $\pt^\part$, where $\pt^{\rm calo}$ and $\pt^{\rm part}$ correspond to the transverse momentum of the reconstructed jet and its matched particle jet, respectively. The jet response distribution is modelled with a Gaussian fit, and its standard deviation is defined as the truth jet $\pt$ resolution. 

\begin{figure}[h!]
\centering
\includegraphics[width=1\linewidth,viewport=35 0 550 325,clip]{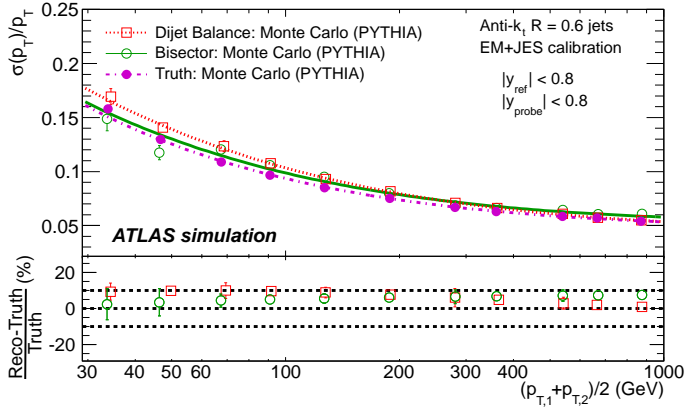}
\caption{Comparison between the Monte Carlo simulation truth jet $\pt$ resolution and the results obtained from the bisector and dijet balance in situ methods (applied to Monte Carlo simulation) for the EM+JES calibration, as a function of $\bar{p}_\T$. The lower panel of the figure shows the relative difference, obtained from the fits, between the in situ methods and Monte Carlo truth results. The  dotted lines indicate a relative difference of $\pm$10\%.
The errors shown are only statistical.
}
\label{fig:ComparisonTruthEMJES}
\end{figure}

The Monte Carlo simulation truth jet $\pt$ resolution is compared to the results obtained from the dijet balance and the bisector in situ methods (applied to Monte Carlo simulation) in Fig.~\ref{fig:ComparisonTruthEMJES}. The agreement between the three sets of points is within 10\%. This result confirms the validity of the physical assumptions discussed in Sections~\ref{sec:dijet} and~\ref{sec:bisector}  and the inference that the observables derived for the in situ MC dijet balance and bisector methods 
provide reliable estimates of the jet energy resolution. The systematic uncertainties on these estimates are of the order of 
10\% (15\%) for jets with $R = 0.6$ ($R = 0.4$), and are discussed in Section~\ref{sec:systematics}.
\section{Jet energy resolution uncertainties}\label{sec:systematics}
\subsection{Experimental uncertainties}\label{sec:SystInsitu}
\begin{figure}[h!]
\centering
\includegraphics[width=1\linewidth,viewport=35 0 550 325,clip]{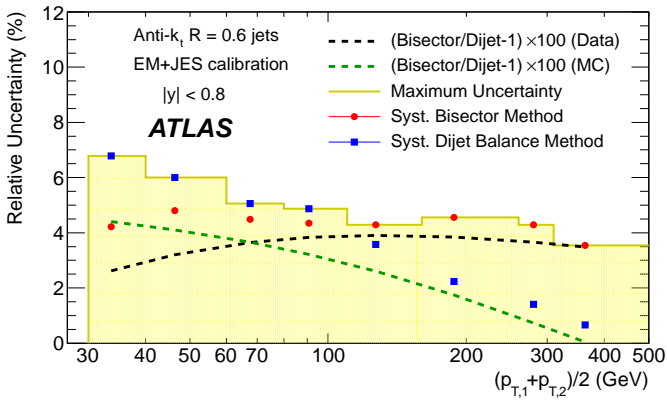}
\caption{The experimental systematic uncertainty on the dijet balance (squares) and bisector (circles) methods as a function of $\bar{p}_\T$, for jets with $|y|<0.8$. The absolute value of the relative difference between the two methods in each $\pt$ bin is also shown for data and for Monte Carlo simulation (dashed lines). 
}
\label{fig:SystInsitu}
\end{figure}

\hspace{\the\parindent}The squares (circles) in Fig.~\ref{fig:SystInsitu} show the experimental relative systematic uncertainty in the dijet balance (bisector) method as a function of $\bar{p}_\T$. The different contributions are discussed below. The shaded area corresponds to the larger of the two systematic uncertainties for each $\bar{p}_\T$ bin. 

For the dijet balance method, systematic uncertainties take into account the variation in resolution when applying different $\Delta\phi$ cuts (varied from 2.6 to 3.0), resulting in a 2--3\% effect for \pt\ = 30--60~GeV, and when varying the soft radiation correction modelling, which contributes up to 6\% at $\pt \approx $ 30~GeV. For the bisector method,  the relative systematic uncertainty is about 4--5\%, and is derived from the precision with which 
the assumption that $\sigma^\part_\psi=\sigma^\part_\eta$ when varying the $p^\mathrm{EM-scale}_{\T,3}$ cut can be verified.  

The contribution from the JES uncertainties~\cite{JESnote} is 1--2\%, determined by re-calculating the jet resolutions after varying the JES within its uncertainty in a fully correlated way. The resolution has also been studied in simulated events with added pile-up events (i.e.\ additional interactions as explained in Section~\ref{sec:PileupMCSample}), as compared to events with one hard interaction only. The sensitivity of the resolution to pile-up is found to be less than $1\%$ for an average number of vertices per event of 1.9.

In summary, the overall relative uncertainty from the in situ methods decreases from about 7\% at $\pt = $30~GeV down to 4\% at $\pt =$ 500~GeV. Figure~\ref{fig:SystInsitu} also shows in dashed lines the absolute value of the relative difference between the two in situ methods, for both data and Monte Carlo simulation. They are found to be in agreement within 4\% up to 500~GeV, and
consistent with these systematic uncertainties.

\subsection{Uncertainties due to the event modelling in the Monte Carlo generators}\label{sec:SystMC}
\begin{figure}[h!]
\begin{center}
\includegraphics[width=1\linewidth,viewport=35 0 550 325,clip]{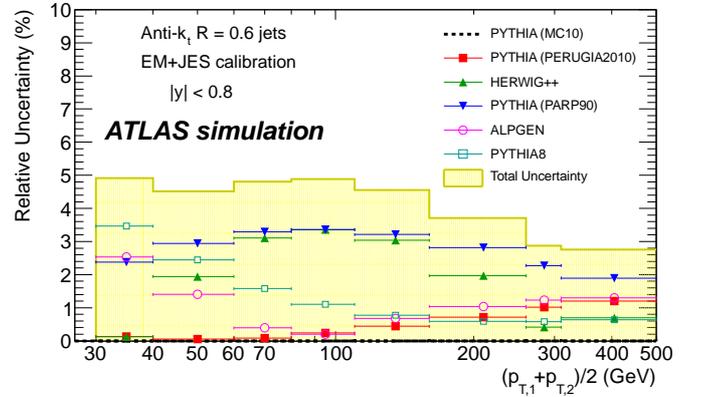} 
\caption{
Systematic uncertainty due to event modelling in Monte Carlo generators on the expected jet energy resolution as a function of $\pt$, for jets with $|y|<0.8$.
The reference is taken from \pythia{} MC10 and other event generators are shown as solid triangles (\herwigpp{}) and open circles (\alpgen{}). Solid squares (\pythia{} {\sc Perugia}\-2010) and inverted triangles (\pythia{} PARP90) summarize differences coming from different tunes and cut-off parameters, respectively. Open squares compare the nominal simulation with \pythia{}8. 
}
\label{fig:SystDiffMC}
\end{center}
\end{figure}

The expected jet \pt\ resolution is calculated for other Monte Carlo simulations in order to assess its dependence on different generator models (\alpgen{} and \herwigpp{}), \pythia{} tunes ({\sc Perugia}\-2010), and other systematic variations (PARP90; see Sec.~\ref{sec:EventGenerators}). Differences between the nominal Monte Carlo simulation and \pythia{}8~\cite{Pythia8} have also been considered. 
These effects, displayed in Fig.~\ref{fig:SystDiffMC}, never exceed 4\%. Although they are not relevant for the in situ measurements of the jet energy resolution themselves, physics analyses sensitive to the expected resolution have to consider a systematic uncertainty from event modelling estimated from the sum in quadrature of the different cases considered here. This is shown by the shaded area in Fig.~\ref{fig:SystDiffMC} and found to be at most 5\%.

\subsection{Uncertainties on the measured resolutions}\label{sec:SystSummary}

\begin{table}[t]
\centering
\begin{tabular}{|c|c|c|c|c| }
\hline 
Jet & Rapidity  &\multicolumn{3}{c|}{Total systematic uncertainty} \\ \cline{3-5}
radius & range &Low \pt&Med \pt&High \pt\\ \hline
\multirow{3}{*}{$R = 0.6$} 
&$\ \ \ 0 \leq |y| < 0.8$ & 12\% & 10\% & 11\% \\ 
&    $0.8 \leq |y| < 1.2$ & 12\% & 10\% & 13\% \\
&    $1.2 \leq |y| < 2.1$ & 14\% & 12\% & 14\% \\
&    $2.1 \leq |y| < 2.8$ & 15\% & 13\% & 18\% \\ \hline
\multirow{3}{*}{$R = 0.4$}           
&$\ \ \ 0 \leq |y| < 0.8$ & 17\% & 15\% & 11\% \\
&    $0.8 \leq |y| < 1.2$ & 20\% & 18\% & 14\% \\
&    $1.2 \leq |y| < 2.1$ & 20\% & 18\% & 14\% \\ 
&    $2.1 \leq |y| < 2.8$ & 20\% & 18\% & 18\% \\
\hline
\end{tabular}
\caption{Relative systematic uncertainties at low ($\sim50$~GeV), medium ($\sim150$~GeV) and high ($\sim400$~GeV) \pt, for the four rapidity regions and the two jet radii studied. The uncertainties are similar for the four calibration schemes.}
\label{tab:sys}
\end{table}
The uncertainties in the measured resolutions are dominated by the systematic uncertainties, which are shown in  Table~\ref{tab:sys} as a percentage of the resolution for the four rapidity regions and the two jet sizes considered, and for  characteristic ranges, low ($\sim$~50~GeV), medium ($\sim150$~GeV) and high ($\sim400$~GeV) \pt. The results are similar for the four calibration schemes.
\begin{figure*}[b!]
\centering
\includegraphics[width=.49\linewidth,viewport=35 0 550 325,clip]{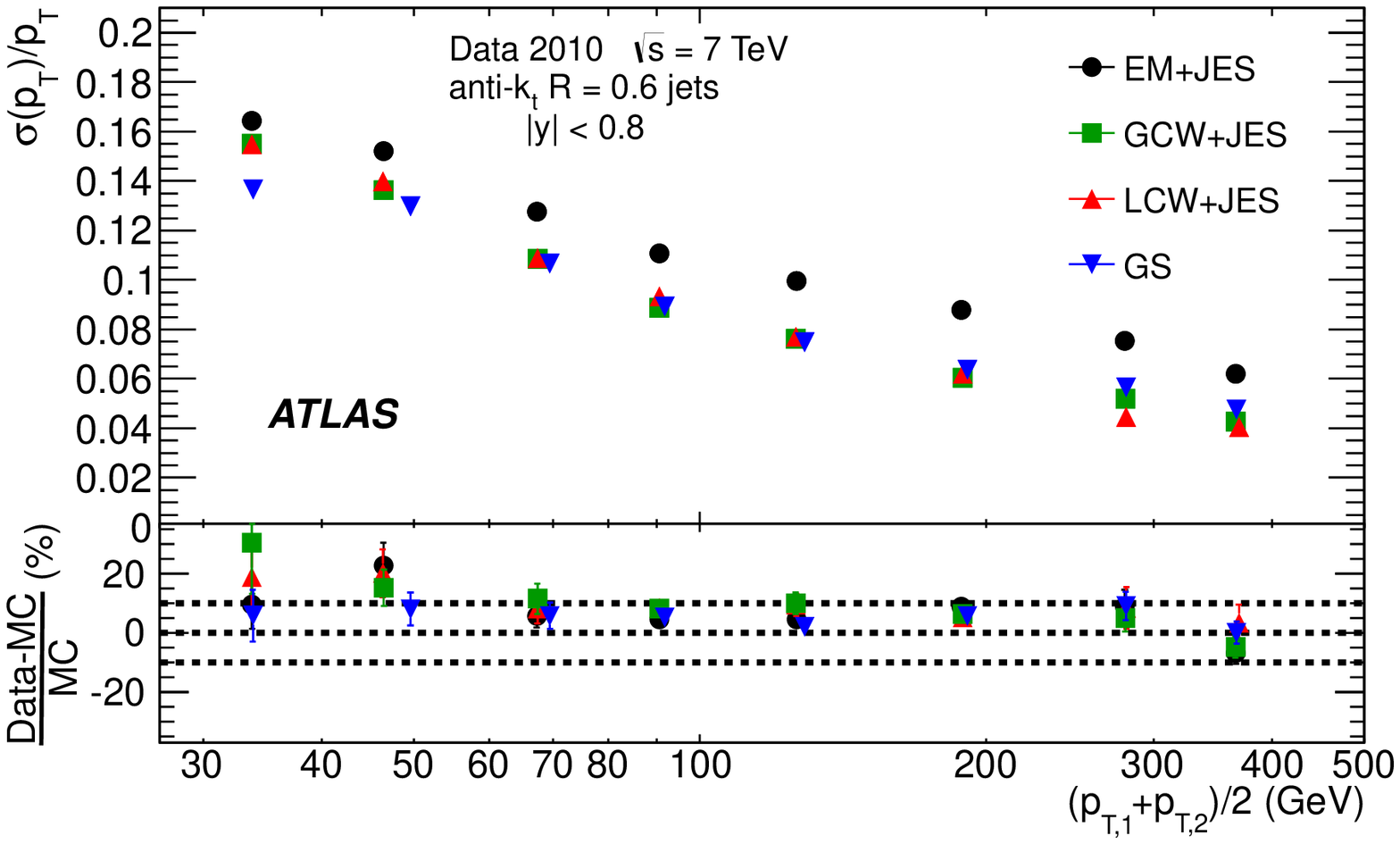}
\includegraphics[width=.49\linewidth,viewport=35 0 550 325,clip]{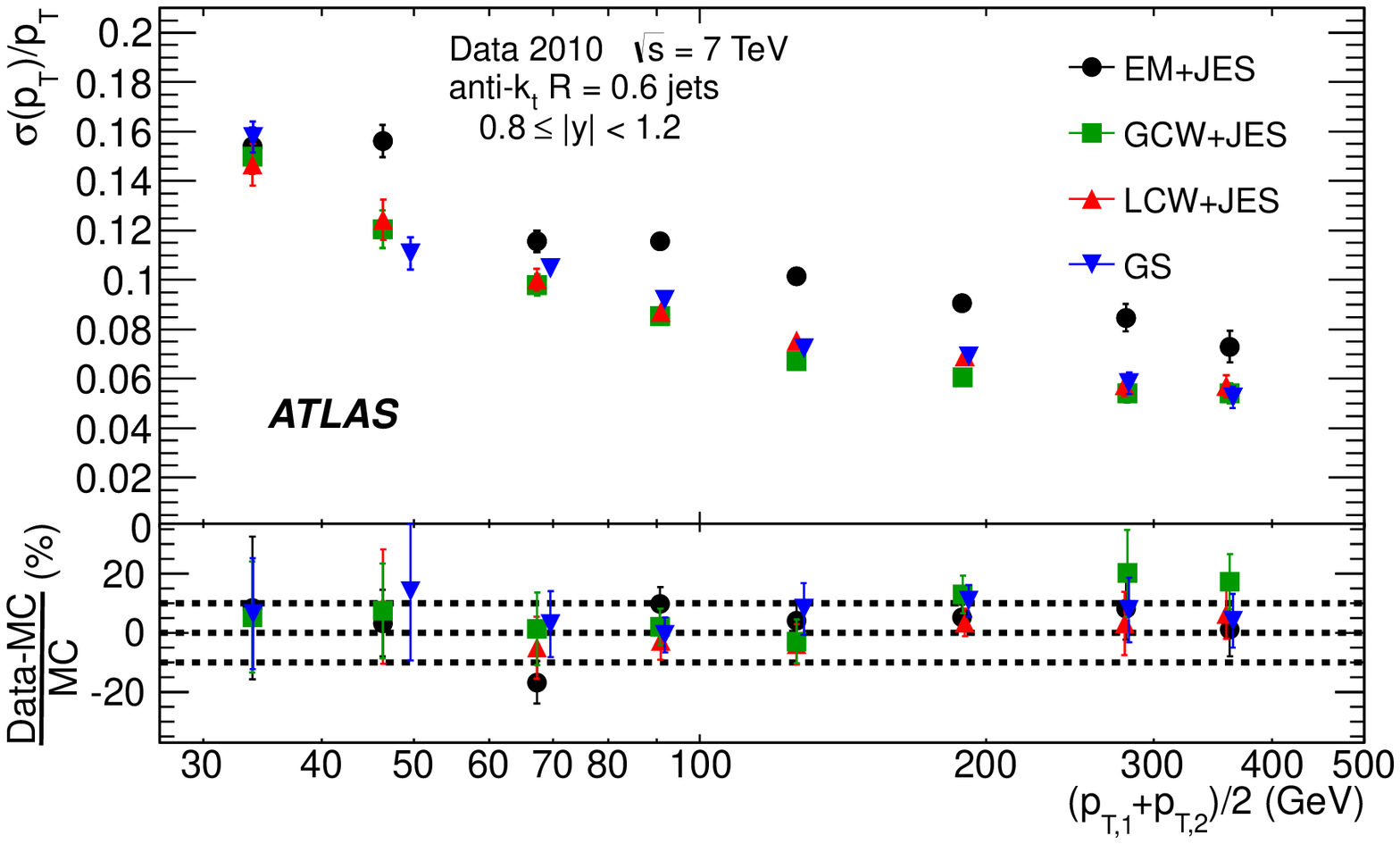}\\
\includegraphics[width=.49\linewidth,viewport=35 0 550 325,clip]{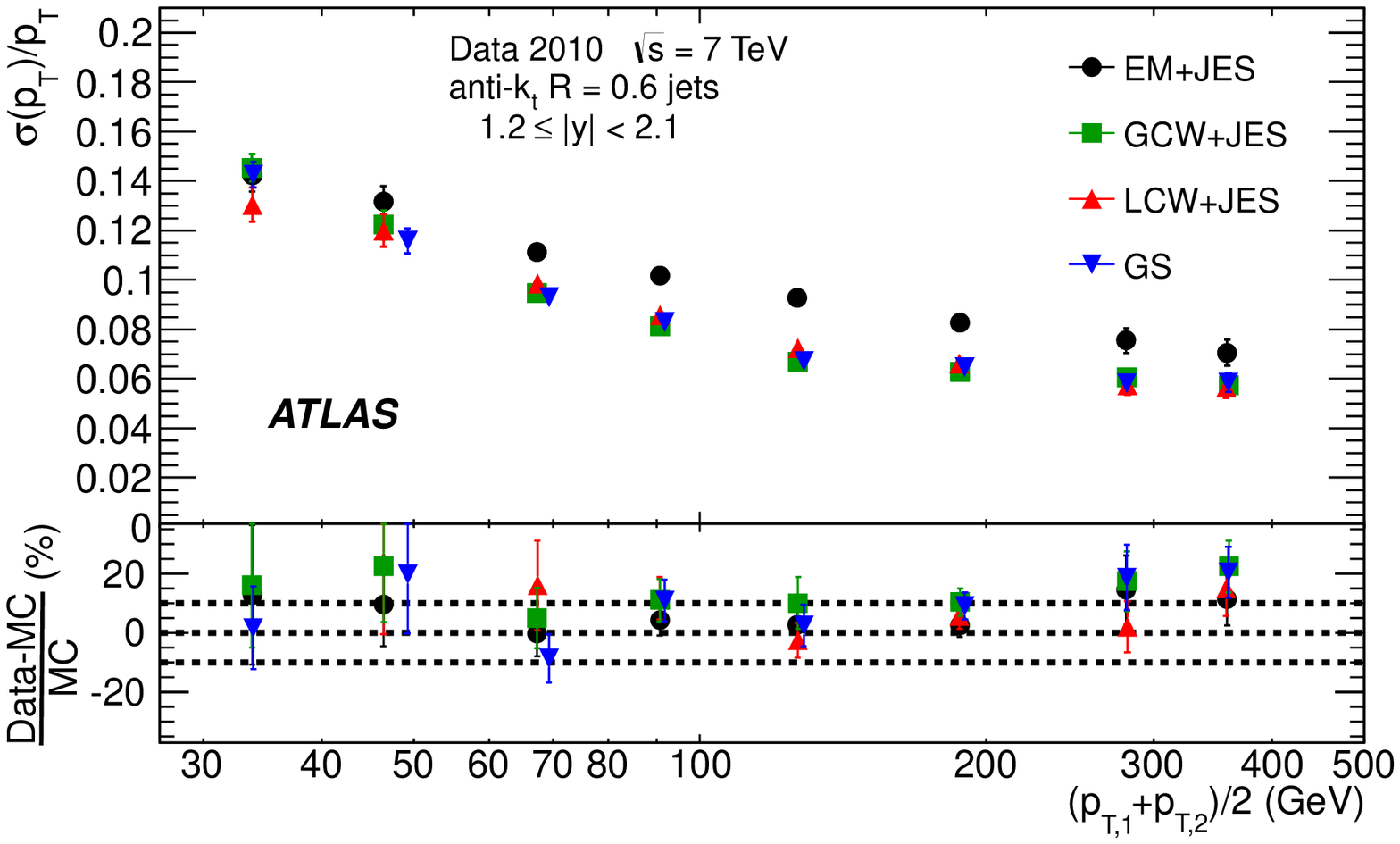}
\includegraphics[width=.49\linewidth,viewport=35 0 550 325,clip]{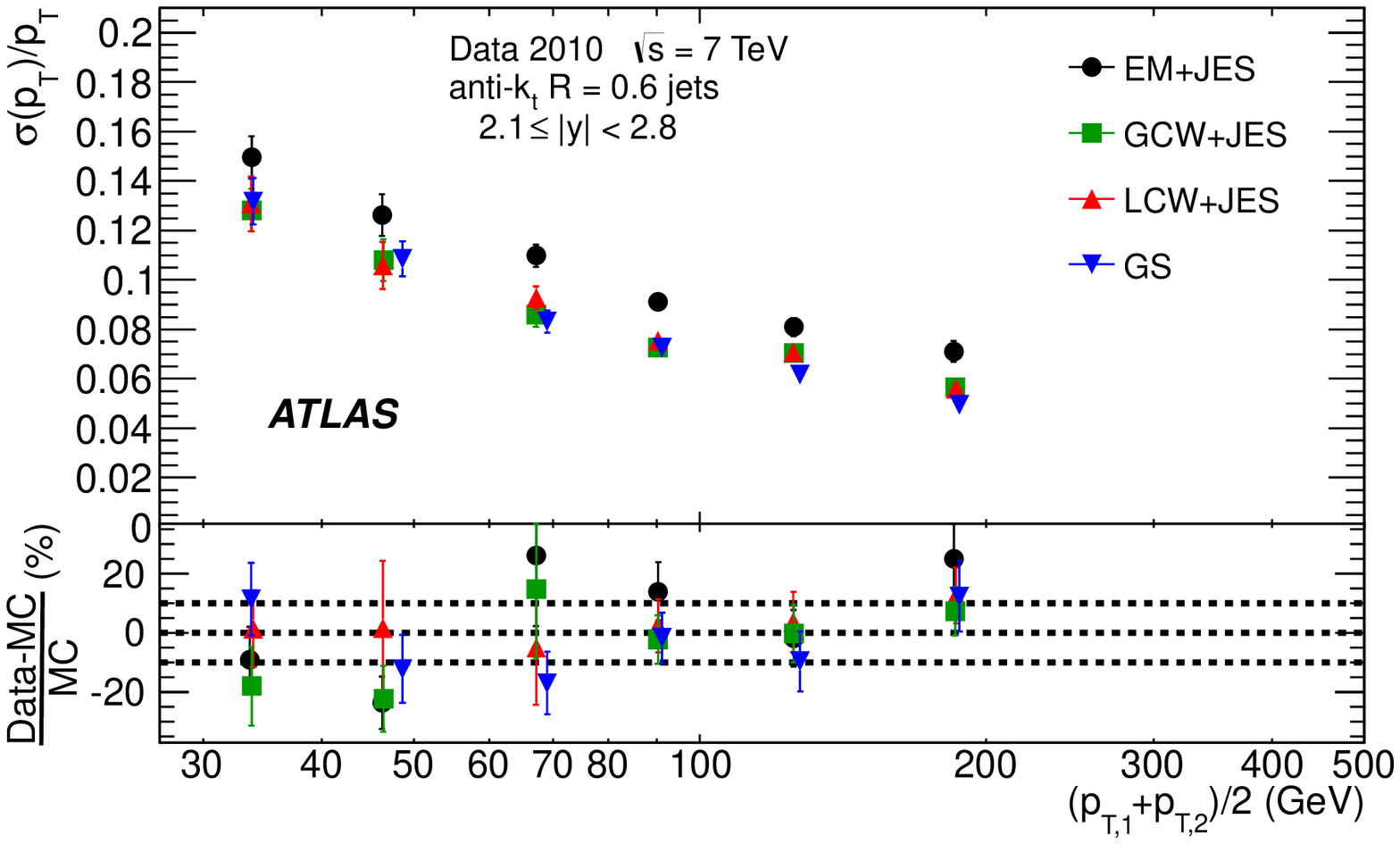}
\caption{
Fractional jet \pt\ resolutions as a function of $\bar{p}_\T$ for anti-$k_t$ jets with $R=0.6$ with $|y| < 0.8$ (top left), $0.8 \leq |y| <  1.2$ (top right), $1.2\leq|y|<2.1$ (bottom left) and $2.1\leq|y|<2.8$ (bottom right), using the bisector in situ method,
for four jet calibration schemes: EM+JES, Local Cluster Weighting (LCW+JES), Global Cell Weighting (GCW+JES) and Global Sequential (GS). The lower panels show the relative difference between data and Monte Carlo simulation results. The dotted lines indicate relative differences of $\pm$10\%.
The errors shown are only statistical.}
\label{fig:DiffJES}
\end{figure*}
The dominant sources of systematic uncertainty are the closure and the data/MC agreement. The closure uncertainty (see Section~\ref{sec:closure}), defined as 
the precision with which in simulation the resolution determined using the in situ method reproduces the truth jet resolution, 
is larger for $R=0.4$ than for $R=0.6$, decreases with \pt, and is basically independent of the rapidity. The data/MC agreement uncertainty is observed to be independent of $R$, larger at low and high \pt\ than at medium \pt, and to grow with rapidity because of the increasingly limited statistical accuracy with which checks can be performed to assess it. Other systematic uncertainties are significantly smaller. They include the validity of the soft radiation hypothesis, the jet energy scale uncertainty and the dependence on the number of pile-up interactions. The uncertainty due to event modelling is not included, as it does not contribute to the in situ measurement itself.

The systematic uncertainties in Table~\ref{tab:sys} for jets with $R=0.4$ are dominated by the contribution from the closure test. They decrease with \pt\ and are constant for the highest three rapidity bins. They are also consistently larger than for the $R=0.6$ case. The systematic uncertainties for jets with $R=0.6$ receive comparable contributions from closure and data/MC agreement. They tend to increase with rapidity and are slightly lower in the medium \pt\ range. The uncertainty increases at high \pt\ for the end-cap, $2.1\leq|y|<2.8$, because of the limited number of events in this region.

\section{Jet energy resolution for other calibration schemes}\label{sec:PerfOther}

\begin{figure*}[b!]
\centering
\begin{tabular}{cc}
\mbox{\epsfxsize=0.48\hsize\epsfbox[25 0 550 325]{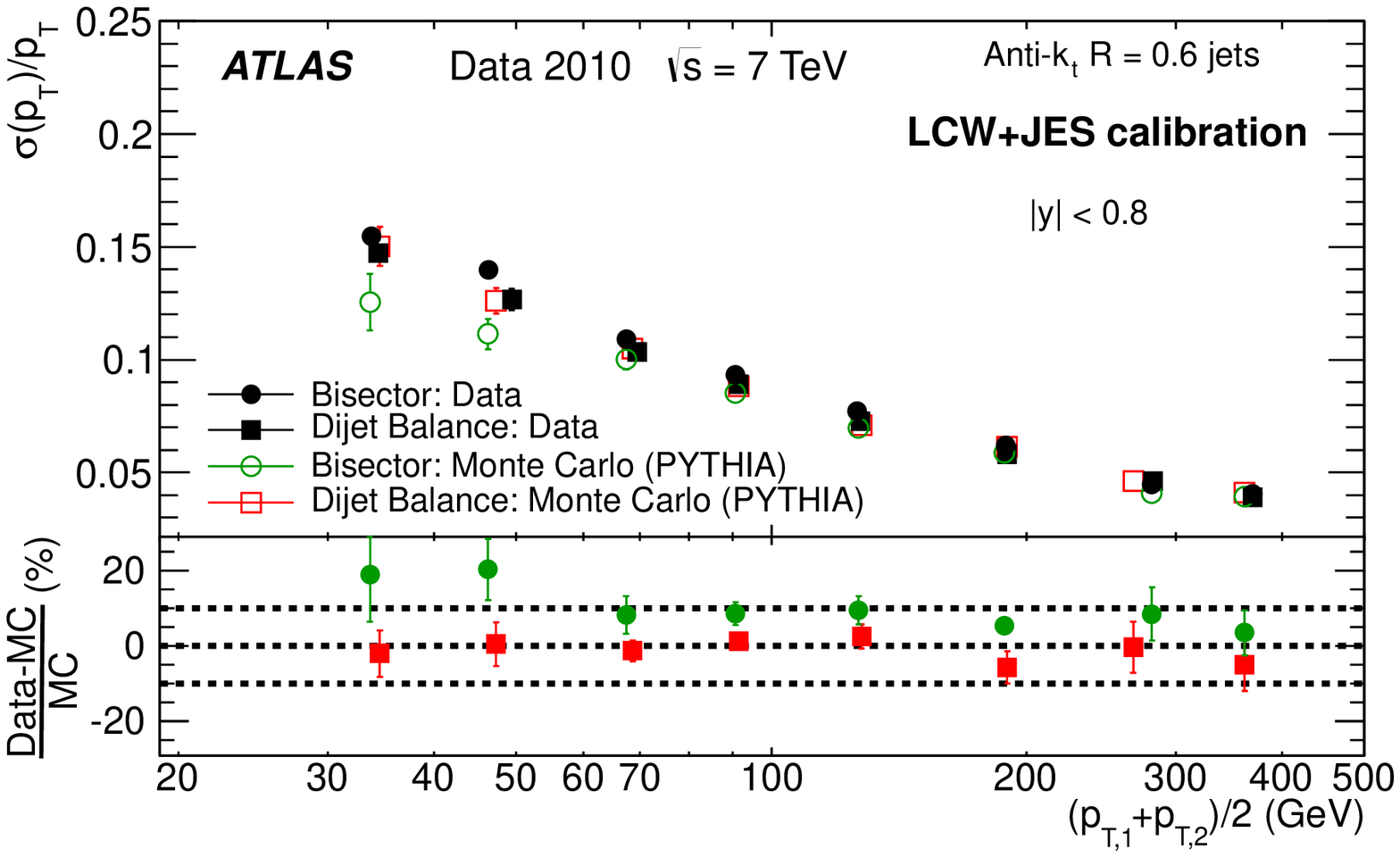}} &
\mbox{\epsfxsize=0.48\hsize\epsfbox[25 0 550 325]{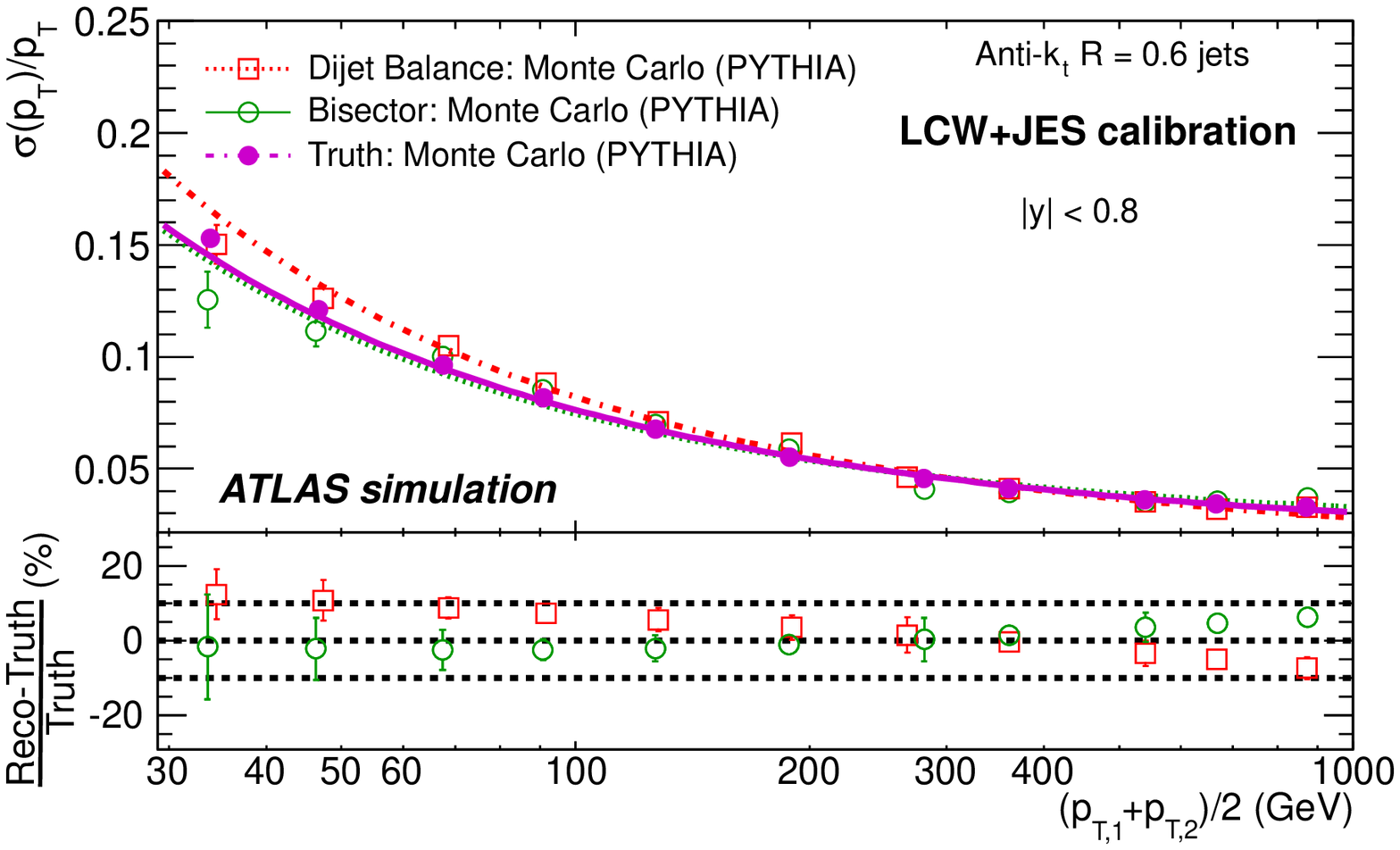}} \\
\mbox{\epsfxsize=0.48\hsize\epsfbox[25 0 550 325]{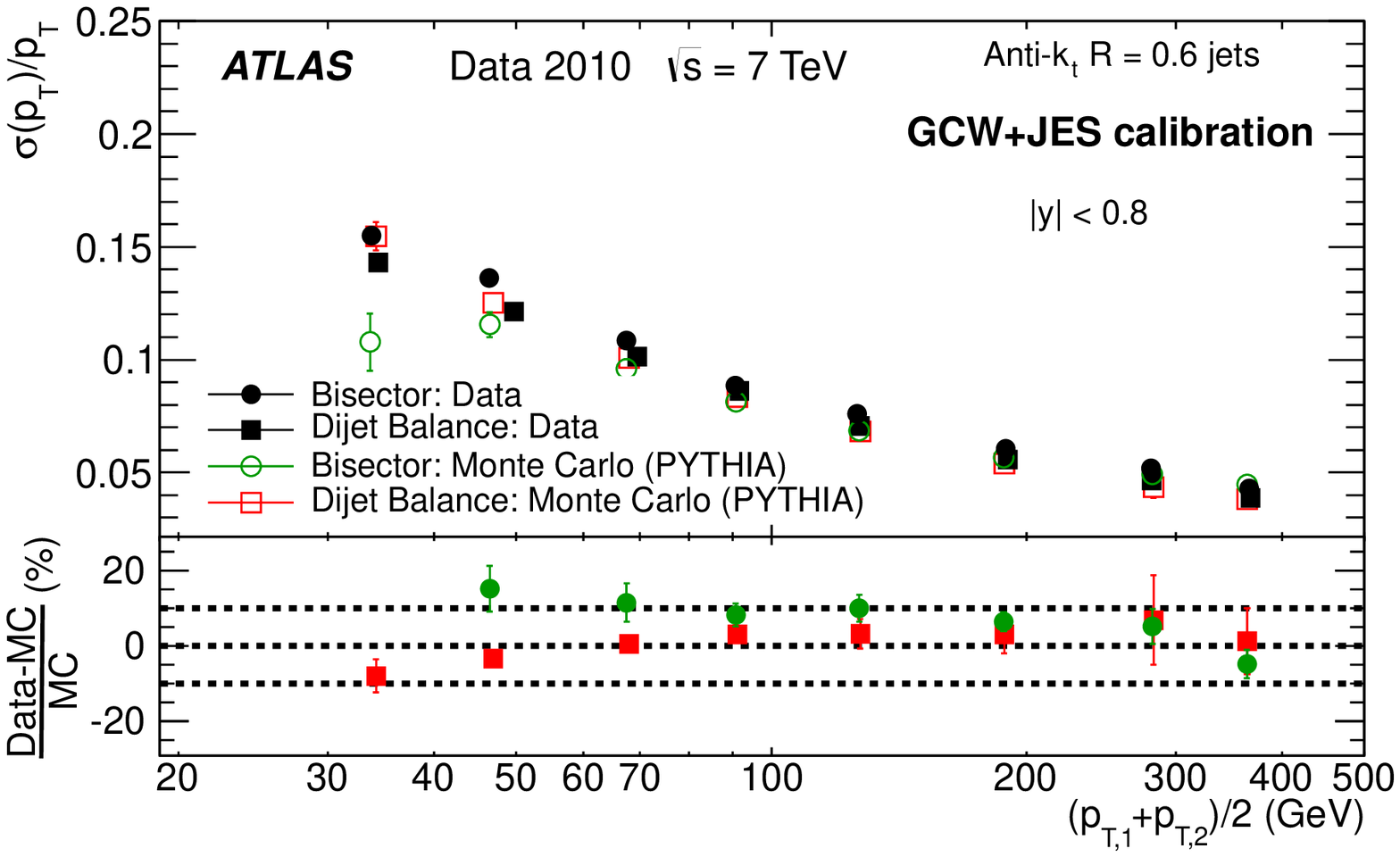}} &
\mbox{\epsfxsize=0.48\hsize\epsfbox[25 0 550 325]{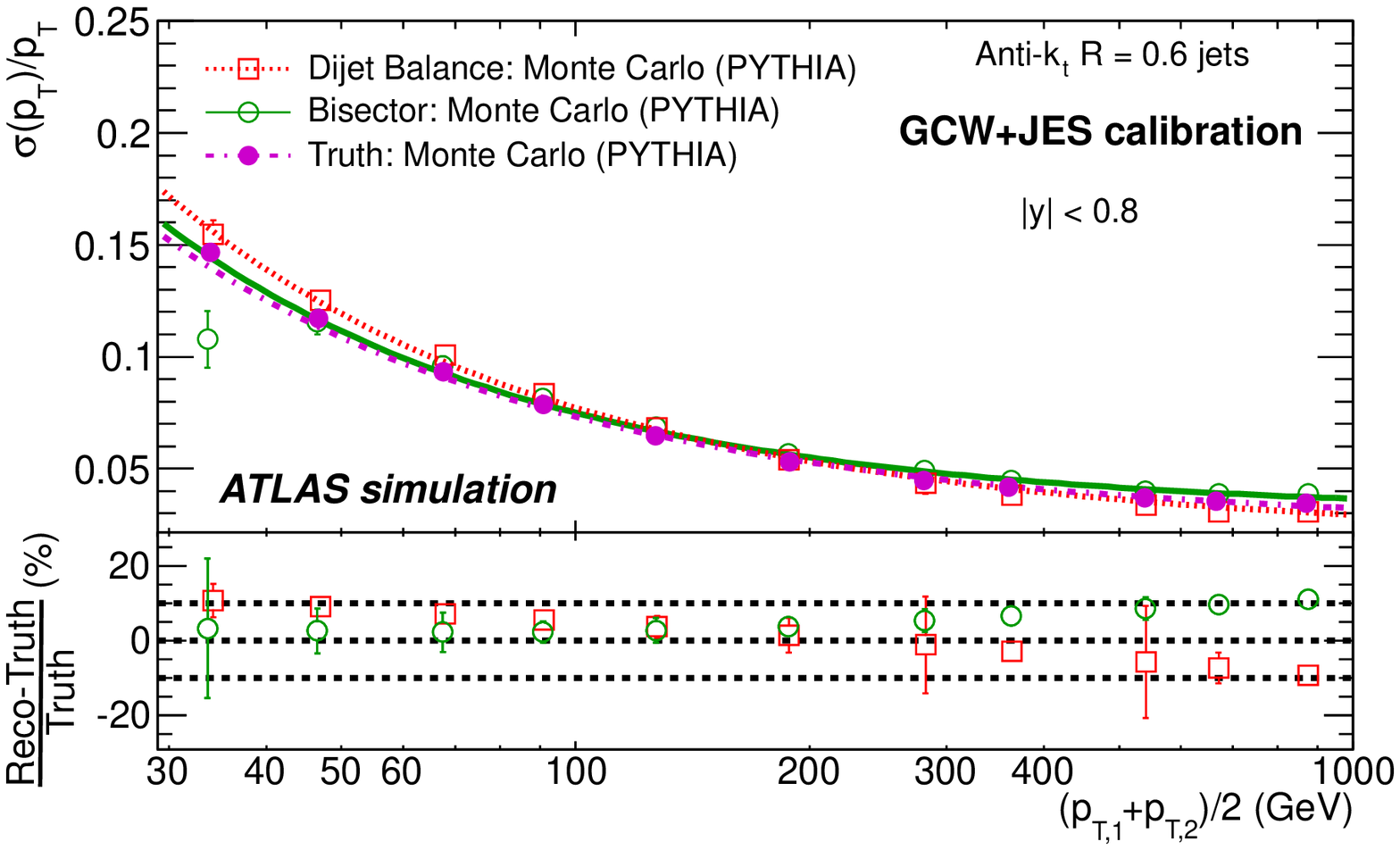}} \\ 
\mbox{\epsfxsize=0.48\hsize\epsfbox[25 0 550 325]{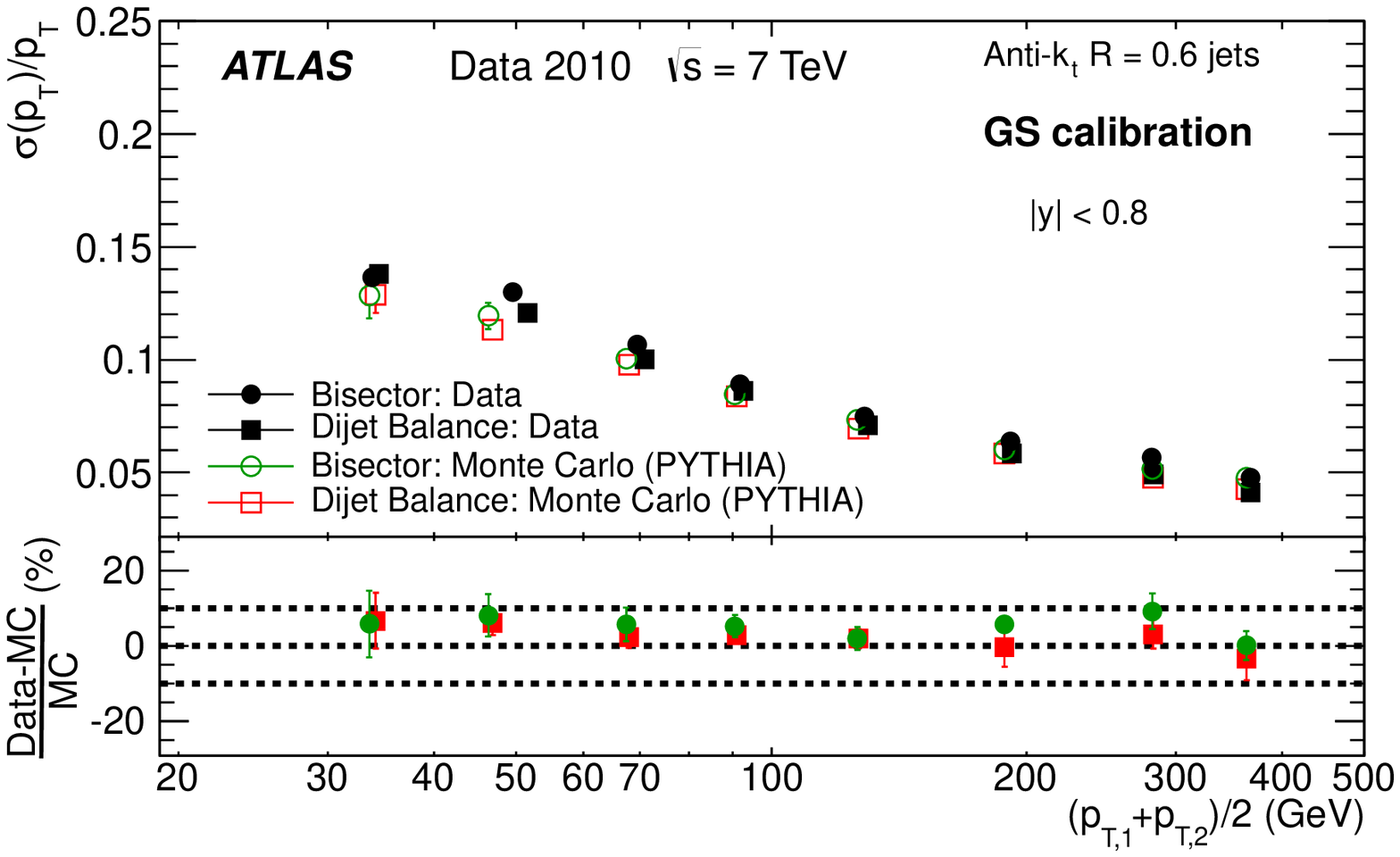}} &
\mbox{\epsfxsize=0.48\hsize\epsfbox[25 0 550 325]{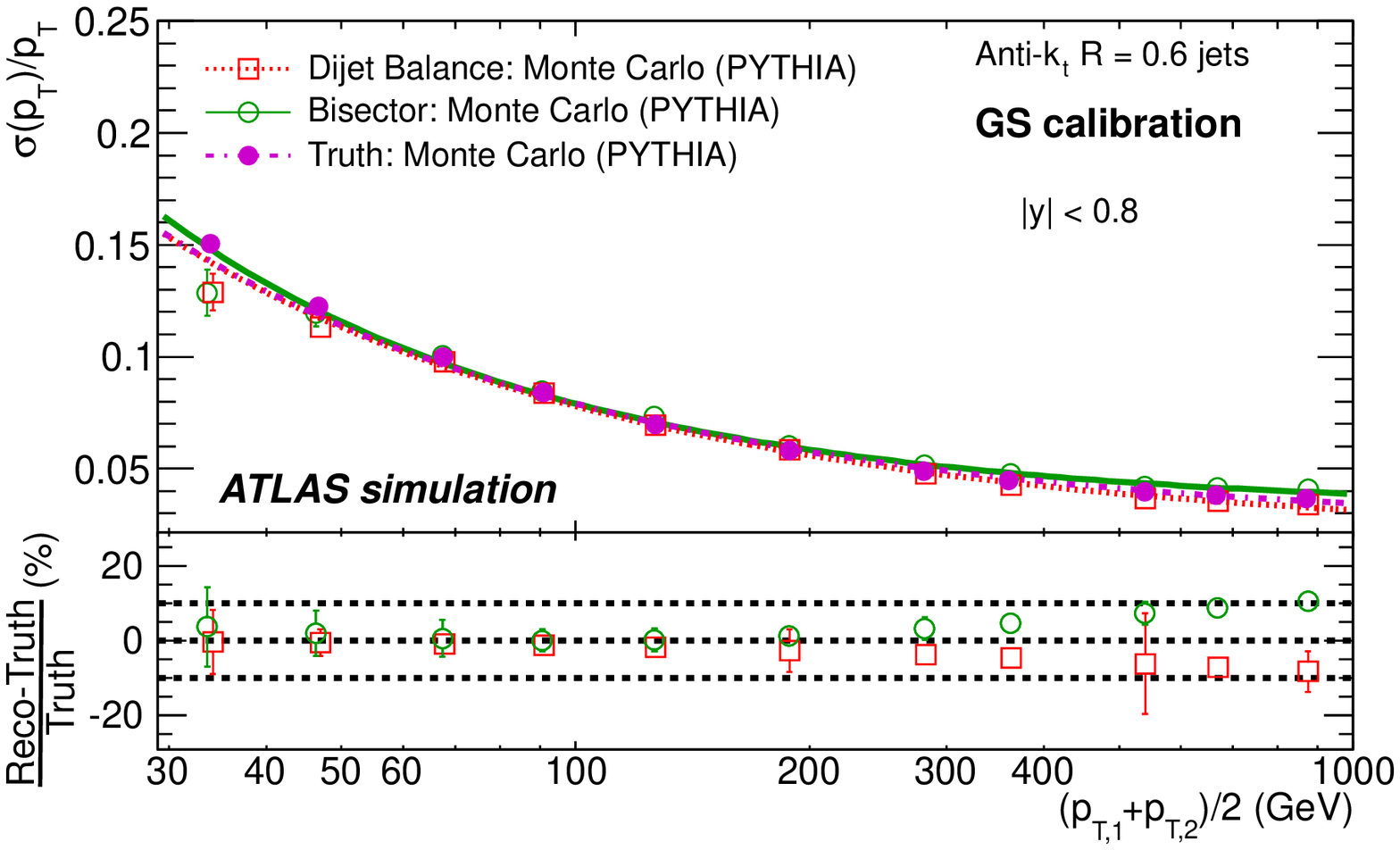}} \\ 
\end{tabular}
\caption{Fractional jet \pt\ resolutions as a function of $\bar{p}_\T$ for anti-$k_t$ jets with $R=0.6$ for the Local Cluster Weighting (LCW+JES), Global Cell Weighting (GCW+JES) and Global Sequential (GS) calibrations.
{\em Left}: Comparison of both in situ methods on data and MC simulation for $|y|<0.8$. 
The lower panels show the relative difference. 
{\em Right}: Comparison between the Monte Carlo simulation truth jet $\pt$ resolution and the final results obtained from the bisector and dijet balance in situ methods (applied to Monte Carlo simulation). The lower panels show the relative differences, obtained from the fits, between the in situ methods and Monte Carlo truth results. 
The dotted lines indicate relative differences of $\pm$10\%.
The errors shown are only statistical.}
\label{fig:ExtraPlots}
\end{figure*}
\hspace{\the\parindent}The resolution performance for anti-$k_t$ jets with $R=0.6$ reconstructed from calo\-rimeter topological clusters for the Local Cluster Weighting (LCW\-+JES), the Global Cell Weighting (GCW+JES) and the Global Sequential (GS) calibration strategies (using the bisector method) is presented in Fig.~\ref{fig:DiffJES} for the Central, Extended Tile Barrel, Transition and End-Cap regions. 
The top part shows the resolutions determined from data, whereas the bottom part compares data and Monte Carlo simulation results. 
The relative improvement in resolution with respect to the EM+JES calibrated jets is comparable for the three more sophisticated calibration techniques. It ranges from 10\% at low $\pt$ up to 40\% at high $\pt$ for all four rapidity regions.

Figure~\ref{fig:ExtraPlots} displays the resolutions for the two in situ methods applied to data and Monte Carlo simulation for $|y|<0.8$ (left plots). It can be observed that the results from the two methods agree, within uncertainties. The Monte Carlo simulation reproduces the data within 10\%.
The figures on the right show the results of a study of the closure for each case, where the truth resolution is compared to that obtained from the in situ methods applied to Monte Carlo simulation data.
The agreement is within 10\%. Overall, comparable agreement in resolution is observed in data and Monte Carlo simulation for the EM+JES, LCW+JES, GCW+JES and GS calibration schemes, with similar systematic uncertainties in the resolutions determined using in situ methods.
\section{Improvement in jet energy resolution using tracks}\label{sec:TBJC}
\hspace{\the\parindent}
The addition of tracking information to the calori\-meter-based energy measurement is expected to compensate for the jet-by-jet fluctuations and improve the jet energy resolution (see Section~\ref{sec:TBJCintro}).
\begin{figure}[b!]
\centering
\includegraphics[width=1\linewidth,viewport=35 0 550 325,clip]{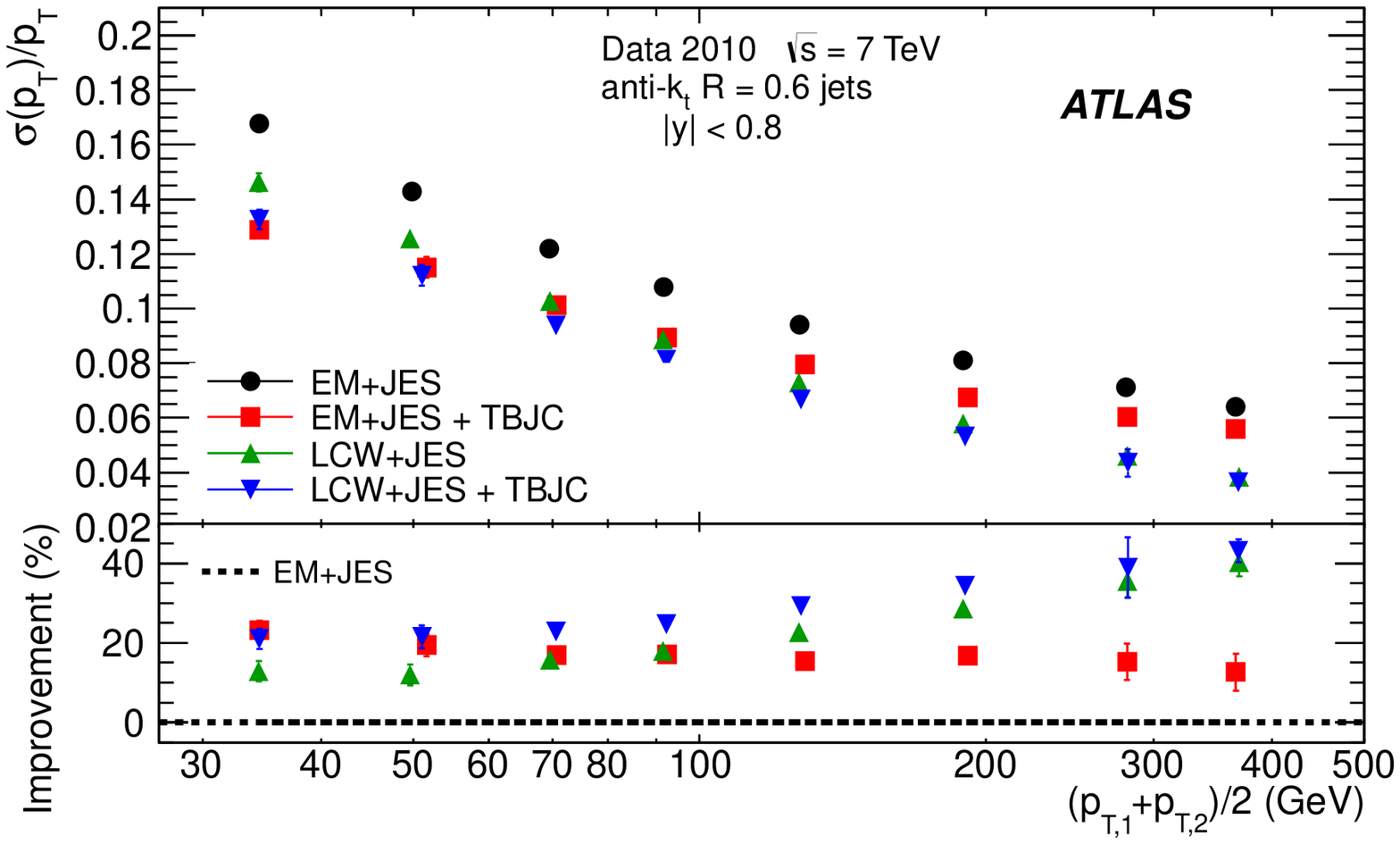}
\includegraphics[width=1\linewidth,viewport=35 0 550 325,clip]{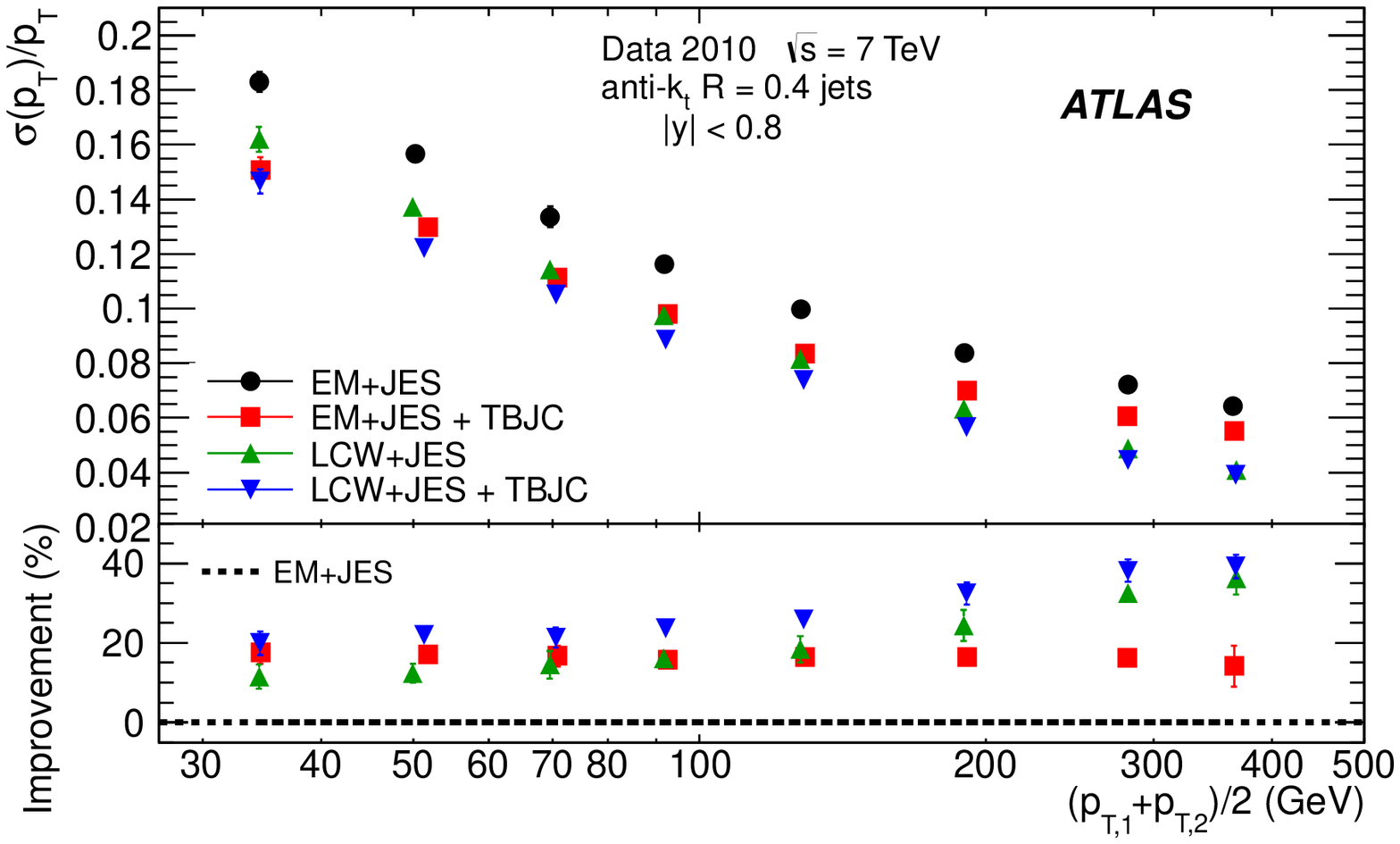}
\caption{
{\em Top}: Fractional jet \pt\ resolutions as a function $\bar{p}_\T$, measured in data for anti-$k_t$ jets with $R=0.6$ (top) and $R=0.4$ (bottom) and for four jet calibration schemes: EM+JES, EM+JES+TBJC, LCW+JES and LCW+JES+TBJC. The lower panel of the figure shows the relative improvement for the EM+JES+TBJC, LCW+JES and LCW+JES+TBJC calibrations with respect to the EM+JES jet calibration scheme, used as reference (dotted line). 
The errors shown are only statistical.}
\label{fig:ComparisonDiffJES}
\end{figure}
The performance of the Track-Based Jet Correction method (TBJC) is studied by applying it to both the EM+JES and LCW+ JES calibration schemes, in the central region. The measured resolution for anti-$k_t$ jets with $R=0.6$ ($R=0.4$) is presented  as a function of the average jet transverse momentum in the top (bottom) plot of Fig.~\ref{fig:ComparisonDiffJES}.

The relative improvement in resolution due to the addition of tracking information is larger at low $\pt$ and more important for the EM+JES calibration scheme. It ranges from 22\% (10\%) at low $\pt$ to 15\% (5\%) at high $\pt$ for the EM+JES (LCW+JES) calibration. 
For $\pt<70$~GeV, jets calibrated with the EM+JES+TBJC scheme show a similar performance to those calibrated with the LCW+JES+TBJC scheme. 
Overall, jets with LCW+JES+TBJC show the best fractional energy resolution over the full \pt\ range.

\section{Summary}\label{sec:Summary}
The jet energy resolution for various JES calibration schemes has been measured using two in situ methods with a data sample corresponding to an integrated luminosity of 35 pb$^{-1}$ collected in 2010 by the \ATLAS{} experiment at $\sqrt{s}$~= 7~TeV. 
%
%
%

The Monte Carlo simulation describes the jet energy resolution measured in data within 10\% for jets with \pt~values between 30~GeV and 500~GeV in the rapidity range $|y|<2.8 $. 

The resolutions obtained applying the in situ techniques to Monte Carlo simulation are in agreement within 10\% with the resolutions determined by comparing jets at calorimeter and particle level.
Overall, the results measured with the two in situ methods have been found to be consistent within systematic uncertainties.

\section{Acknowledgements}\label{sec:Ack}

We thank CERN for the very successful operation of the LHC, as well as the
support staff from our institutions without whom ATLAS could not be
operated efficiently.

We acknowledge the support of ANPCyT, Argentina; YerPhI, Armenia; ARC,
Australia; BMWF and FWF, Austria; ANAS, Azerbaijan; SSTC, Belarus; CNPq and FAPESP,
Brazil; NSERC, NRC and CFI, Canada; CERN; CONICYT, Chile; CAS, MOST and NSFC,
China; COLCIENCIAS, Colombia; MSMT CR, MPO CR and VSC CR, Czech Republic;
DNRF, DNSRC and Lundbeck Foundation, Denmark; EPLANET and ERC, European Union;
IN2P3-CNRS, CEA-DSM/IRFU, France; GNSF, Georgia; BMBF, DFG, HGF, MPG and AvH
Foundation, Germany; GSRT, Greece; ISF, MINERVA, GIF, DIP and Benoziyo Center,
Israel; INFN, Italy; MEXT and JSPS, Japan; CNRST, Morocco; FOM and NWO,
Netherlands; BRF and RCN, Norway; MNiSW, Poland; GRICES and FCT, Portugal; MERYS
(MECTS), Romania; MES of Russia and ROSATOM, Russian Federation; JINR; MSTD,
Serbia; MSSR, Slovakia; ARRS and MVZT, Slovenia; DST/NRF, South Africa;
MICINN, Spain; SRC and Wallenberg Foundation, Sweden; SER, SNSF and Cantons of
Bern and Geneva, Switzerland; NSC, Taiwan; TAEK, Turkey; STFC, the Royal
Society and Leverhulme Trust, United Kingdom; DOE and NSF, United States of
America.

The crucial computing support from all WLCG partners is acknowledged
gratefully, in particular from CERN and the ATLAS Tier-1 facilities at
TRIUMF (Canada), NDGF (Denmark, Norway, Sweden), CC-IN2P3 (France),
KIT/GridKA (Germany), INFN-CNAF (Italy), NL-T1 (Netherlands), PIC (Spain),
ASGC (Taiwan), RAL (UK) and BNL (USA) and in the Tier-2 facilities
worldwide.

\bibliographystyle{atlasnote}
\bibliography{JERPaper2010}

\onecolumn 
\clearpage 
 
\begin{flushleft}
{\Large The ATLAS Collaboration}

\bigskip

G.~Aad$^{\rm 47}$,
T.~Abajyan$^{\rm 20}$,
B.~Abbott$^{\rm 110}$,
J.~Abdallah$^{\rm 11}$,
S.~Abdel~Khalek$^{\rm 114}$,
A.A.~Abdelalim$^{\rm 48}$,
O.~Abdinov$^{\rm 10}$,
R.~Aben$^{\rm 104}$,
B.~Abi$^{\rm 111}$,
M.~Abolins$^{\rm 87}$
% ...
\bigskip

% ...
$^{10}$ Institute of Physics, Azerbaijan Academy of Sciences, Baku, Azerbaijan\\
$^{11}$ Institut de F\'{i}sica d'Altes Energies and Departament de F\'{i}sica de la Universitat Aut\`{o}noma de Barcelona and ICREA, Barcelona, Spain\\
% ...
$^{20}$ Physikalisches Institut, University of Bonn, Bonn, Germany\\
% ...
$^{47}$ Fakult\"{a}t f\"{u}r Mathematik und Physik, Albert-Ludwigs-Universit\"{a}t, Freiburg, Germany\\
$^{48}$ Section de Physique, Universit\'{e} de Gen\`{e}ve, Geneva, Switzerland\\
% ...
$^{87}$ Department of Physics and Astronomy, Michigan State University, East Lansing MI, United States of America\\
% ...
$^{104}$ Nikhef National Institute for Subatomic Physics and University of Amsterdam, Amsterdam, Netherlands\\
% ...
$^{110}$ Homer L. Dodge Department of Physics and Astronomy, University of Oklahoma, Norman OK, United States of America\\
$^{111}$ Department of Physics, Oklahoma State University, Stillwater OK, United States of America\\
% ...
$^{114}$ LAL, Universit\'{e} Paris-Sud and CNRS/IN2P3, Orsay, France\\
% ...
\end{flushleft}

\end{document}